\documentclass[preprint]{aastex}

\shorttitle{2MASS Ultracool Dwarfs} \shortauthors{Cruz et al.}

\begin{document}

\title{Meeting the Cool New Neighbors V:  A 2MASS-Selected Sample of Ultracool
Dwarfs}

\author{Kelle L. Cruz\altaffilmark{1}}
\affil{Department of Physics and Astronomy, University of Pennsylvania,
209 South 33rd Street, Philadelphia, PA 19104; kelle@sas.upenn.edu}

\author{I. Neill Reid\altaffilmark{1}}
\affil{Space Telescope Science Institute, 3700 San Martin Drive, Baltimore, MD 21218; and
Department of Physics and Astronomy, University of Pennsylvania, 209 South 33rd Street,
Philadelphia, PA  19104; inr@stsci.edu}

\author{James Liebert\altaffilmark{1}}
\affil{Department of Astronomy and Steward Observatory,
University of Arizona, Tucson, AZ 85721; liebert@as.arizona.edu}

\and

\author{J. Davy Kirkpatrick, Patrick J. Lowrance\altaffilmark{1}}
\affil{Infrared Processing and Analysis Center, California
Institute of Technology, Pasadena, CA 91125;
davy@ipac.caltech.edu, lowrance@ipac.caltech.edu}

\altaffiltext{1}{Visiting Astronomer, Kitt Peak National
Observatory, NOAO, which is operated by AURA under cooperative
agreement with the NSF.}

\begin{abstract}
We present initial results of our effort to create a statistically robust,
volume-limited sample of ultracool dwarfs from the 2MASS Second
Incremental Data Release.  We are engaged in a multifaceted search for
nearby late-type dwarfs and this is the first installment of our search
using purely photometric selection. The goal of this work is a
determination of the low-mass star and brown dwarf luminosity function in
the infrared. Here, we outline the construction of the sample, dubbed
2MU2, and present our first results, including the discovery of 186 M7--L6
dwarfs---47 of these are likely to be within 20 pc of the Sun. These
results represent 66\% of the ultracool candidates in our sample yet
constitute an 127\% increase in the number of ultracool dwarfs known
within the volume searched (covering 40\% of the sky out to 20 pc).  In
addition, we have identified 10 M4--M6.5 objects that are likely to be
within 20 pc (or within 1$\sigma$). Finally, based on these initial data,
we present a preliminary luminosity function and discuss several
interesting features of the partial sample presented here.  Once our
sample is complete, we will use our measured luminosity function to
constrain the mass function of low-mass stars and brown dwarfs.
\end{abstract}


\section{Introduction}

Ultracool dwarfs (spectral types M7 and later) include stars and
brown dwarfs and can have masses as small as several Jupiter
masses.  The search for these ultracool dwarfs has been greatly
enabled by three recent deep, wide-sky surveys:  the Two-Micron
All Sky Survey \citep[2MASS]{2MASS}, the Deep Near Infrared
Southern Sky survey \citep[DENIS]{DENIS}, and the Sloan Digital
Sky Survey \citep[SDSS]{york}.

Two methods are used to identify late-type dwarfs in these large
data sets: direct query of the survey's photometric data
products; and using those data in combination with new and
existing proper motion data and optical photometry.  The former
method has yielded close to 200 L dwarfs and over 180 ultracool M
dwarfs in the three surveys (notable for quantity are \citet{NN5,
NN4, SDSS2, Hawley, NN, K00, K99, M99}). The other tack has
unveiled many previously unknown late-type members of our Solar
neighborhood (if not as many L dwarfs). Near-infrared photometry
from DENIS and 2MASS has been combined with Luyten's proper
motion catalogs (\citet[LHS]{lhs}, \citet[NLTT]{nltt}), new
proper-motion catalogs \citep{LSR}, and with new and existing
optical photometry to identify nearly 200 M dwarfs but less than
10 L dwarfs within 35 pc of the Sun \citep{LSR2, LSR1, Reyle,
Lodieu, Scholz, paper1, paper2, paper3, EROS}.

While both methods have yielded many discoveries, both have biases and
limitations that make it difficult to use these data sets to study the new
population of ultracool dwarfs statistically.  The use of proper motion
and optical data has the advantage of finding the nearest objects but this
type of study is restricted to bright magnitudes and hence, earlier-type
objects. In addition, proper motion searches are strongly biased towards
objects with intrinsically high space motions with respect to the Sun.
Photometric searches avoid a kinematic bias and probe much larger spatial
volumes by reaching fainter magnitudes than proper motion surveys.
However, previous photometric searches are characterized by small sky
coverage and/or restrictive color-criteria.

Our aim is to use both proper motion and photometric searches to complete
the census of the nearest ultracool dwarfs.  In the first three papers in
this series \citep[hereafter Papers I, II, \& III]{paper1, paper2, paper3}
we presented the first results of our effort to cross-reference the 2MASS
Second Incremental Release with the NLTT catalog.  While we uncovered
nearly 150 new objects within 20 pc, most of them are early and mid-M
dwarfs. \citet{paper4} (hereafter Paper IV) presented the discovery of an
M8.5 dwarf within 6 pc found in the course of the program described in
this Paper.  Here, we present the initial results and the methodology of
our photometric search of the 2MASS Second Incremental Release for nearby
objects cooler than spectral type M6.

Building on this foundation, we have created a statistically robust sample
of ultracool dwarfs from the 2MASS Second Incremental Release that is
complete within 20 pc of the Sun for spectral types M7 to L8.  The sample
is dubbed 2MU2---2MASS ultracool dwarfs from the Second Data Release. This
volume-limited sample permits detailed investigation of the overall range
of properties of these low-mass objects which span the stellar/brown dwarf
boundary. In particular, once the observations are complete, we will have
an infrared luminosity function which we will use to constrain the mass
function of low-mass stars and brown dwarfs.

The current knowledge of the field mass function of low-mass stars
is based primarily on two projects:  the Palomar/Michigan State
University Nearby-Star Spectroscopic Survey \citep{PMSU,PMSU4};
and the 8 pc sample \citep{rg97,8pc,paper4} (hereafter PMSU and 8
pc sample).  Both datasets yield a mass function that is
consistent with a power-law distribution ($\Psi(M)=M^{-\alpha}$)
with $\alpha=1.2$ at low masses ($M<0.6M_{\sun}$) (see Figure 12
of \citet{PMSU4}). However, both samples only extend to
$0.08M_{\sun}$---just above the stellar/brown dwarf boundary for
solar metallicity objects. \citet{8pc} made the first attempt to
extend coverage to substellar mass objects in the field, but their
study is hampered by sparse statistics.  The project described in
this Paper is the first concerted effort focused on studying the
field mass function in the regime of low-mass stars and brown
dwarfs.

This is the fifth paper in a series that report results of our
multifaceted study of nearby late-type objects in the immediate Solar
neighborhood. While the previous papers in the series have concentrated on
objects with large proper motions, here we present our first results of
purely photometric selection of candidate nearby dwarfs using the 2MASS
survey. In \S \ \ref{sec:sample} we describe the creation of the 2MU2
sample. We present our observations in \S \ \ref{sec:obs} and the spectral
type and absolute magnitude calibrations in \S\ \ref{sec:results}. We
discuss interesting objects in \S \ \ref{sec:obj} and present the
characteristics of the portion of the 2MU2 sample presented here and a
preliminary luminosity function in \S \ \ref{sec:lf}. Our conclusions are
in \S \ \ref{sec:conclusions}.

\section{The 2MU2 Sample}
\label{sec:sample} The primary goal of this project is to create a
sample of ultracool dwarfs within 20 pc using the 2MASS Second
Incremental Data Release.  This catalog covers 48\% of the sky and
contains over 162 million point sources with $J$ (1.25
$\micron$), $H$ (1.65 $\micron$), and $K_S$ (2.17 $\micron$)
photometry and highly accurate astrometry. While we use several
methods to refine the 2MU2 sample, as described in detail below,
there are three primary selection criteria. The first requires
targets to have a galactic latitude greater than $10\degr$ in
order to avoid the Galactic plane. Two additional cuts are based
on the $JHK_S$ color-color and color-magnitude sequences of M and
L dwarfs with trigonometric parallax measurements. The final 2MU2
sample of viable candidates contains 1225 targets and covers 40\%
(16,350 sq. deg.) of the sky.  Full details of the selection
procedures are given in this section.

\subsection{Selection Criteria}
\label{sec:cuts} The initial 2MASS query, using the Gator tool
provided by IRSA\footnote{\url{http://irsa.ipac.caltech.edu/}},
required $|b|>10\degr$, $J-K_S>1$, and rejected extended objects.
These criteria selected $\sim$11.3 million sources. Custom built
IDL code was used to further cut the sample in a series of steps
which are detailed below and summarized in Table \ref{tab:cuts}.

\begin{figure}
\epsscale{1.0} \plotone{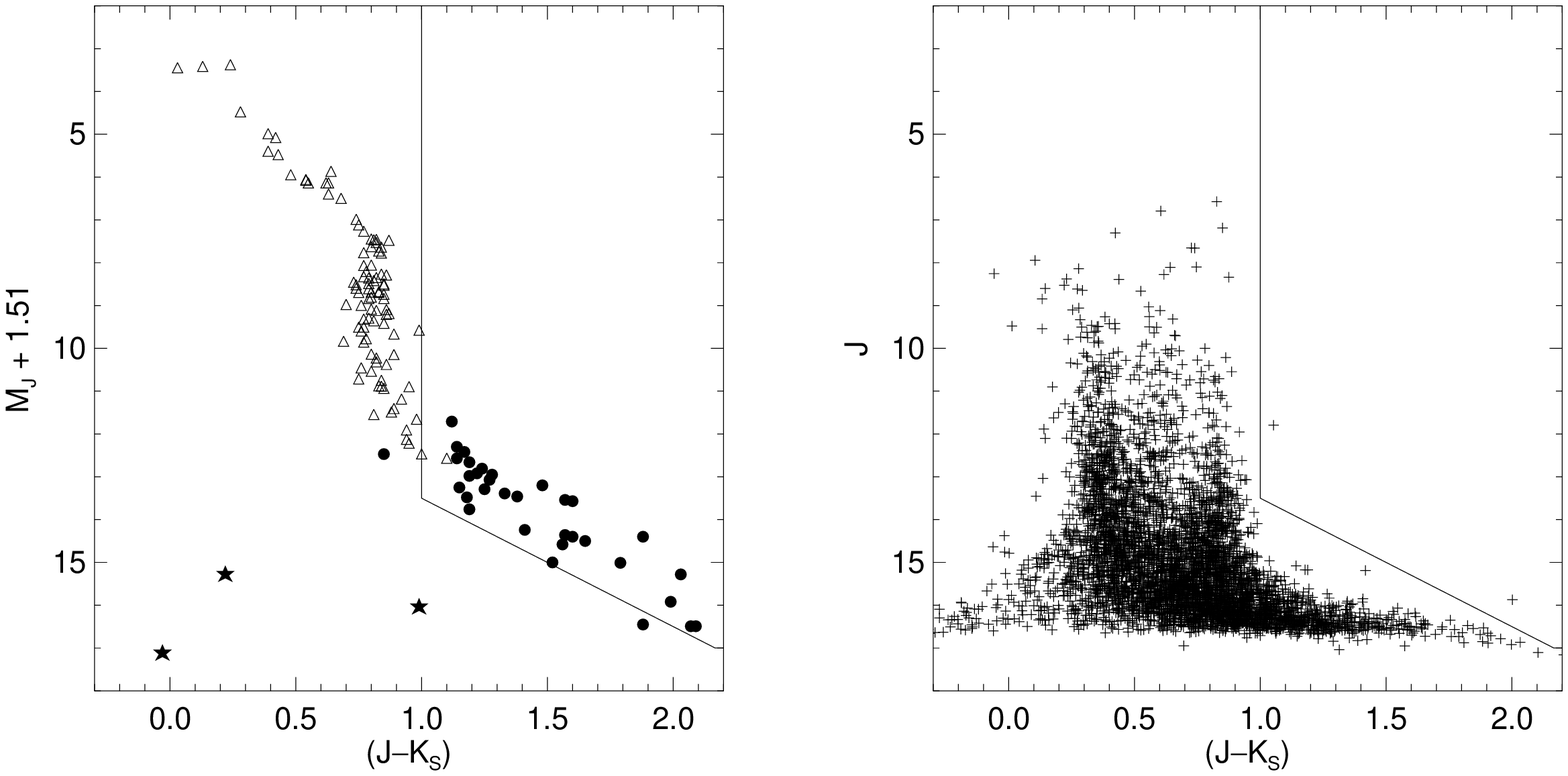}

\caption{Color-magnitude diagram for low-mass stars with trigonometric
parallax measurements shifted to 20 pc (\emph{left panel}) and a typical
$1\degr$ 2MASS field (\emph{right panel}) with our selection criteria.
Triangles are from the 8 pc sample.  Data for ultracool dwarfs (M7--L8,
filled circles) and T dwarfs (filled five-point stars) are from
\citet{Dahn}. Targets are selected if they lie above and to the right of
the cuts.} \label{fig:fig1}
\end{figure}

\begin{figure}

\epsscale{1.0} \plotone{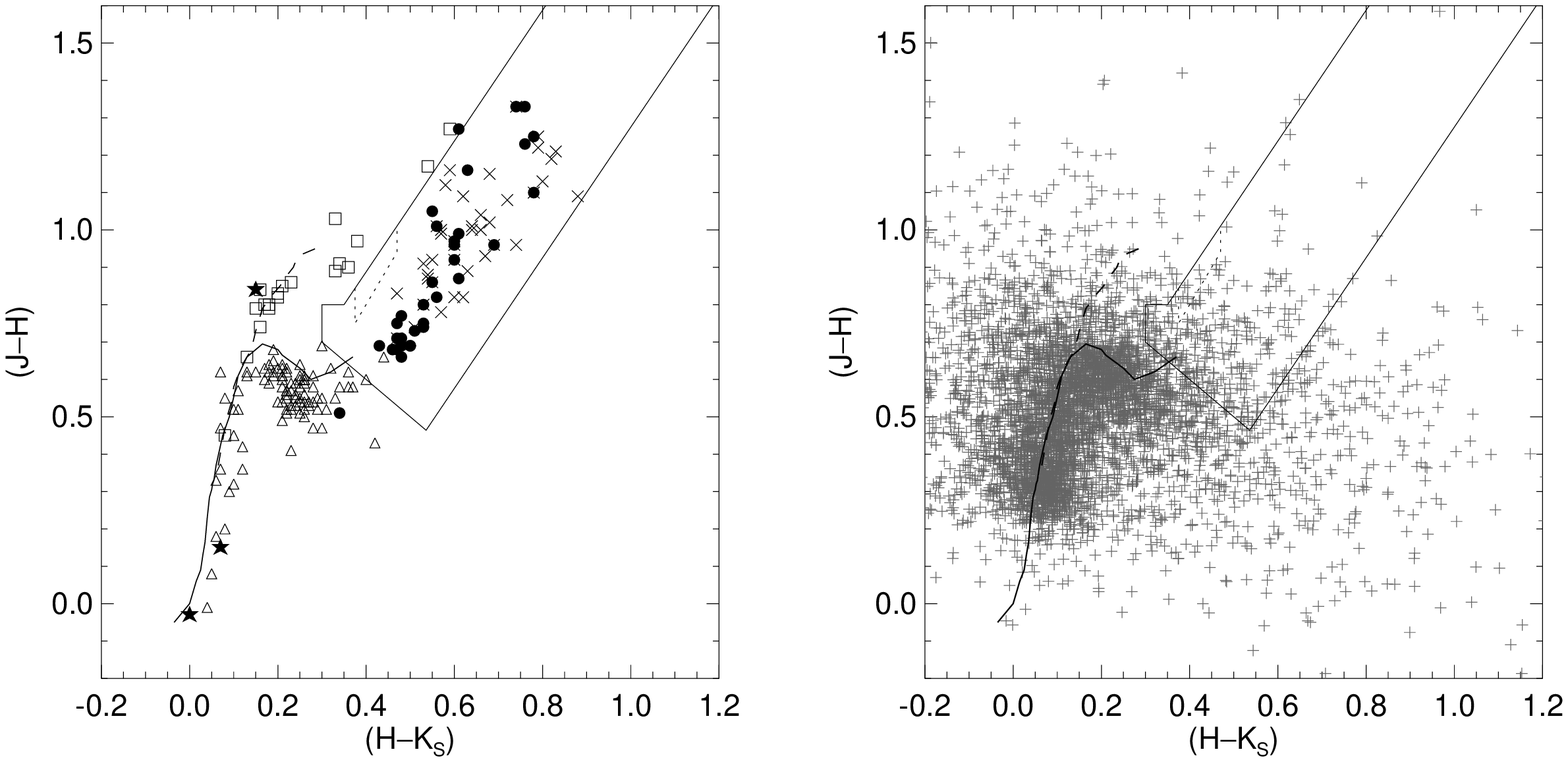}

\caption{Color-color diagram for the same data (plotted with the same
symbols) as Figure 1 and our selection criteria.  In addition to objects
with trigonometric parallax data, we show L dwarfs from \citet{K99,K00}
with uncertainties less than 0.1 magnitudes as crosses and open squares
are giants. We also show the dwarf and giant sequences (both from
\citet{bb}, transformed to the 2MASS system).  Targets are selected if
they lie within the enclosed region.  The region where bright objects
($J<10$) are eliminated as giants is enclosed with a dotted line.}
\label{fig:fig2}
\end{figure}

\begin{figure}
\epsscale{1.0} \plotone{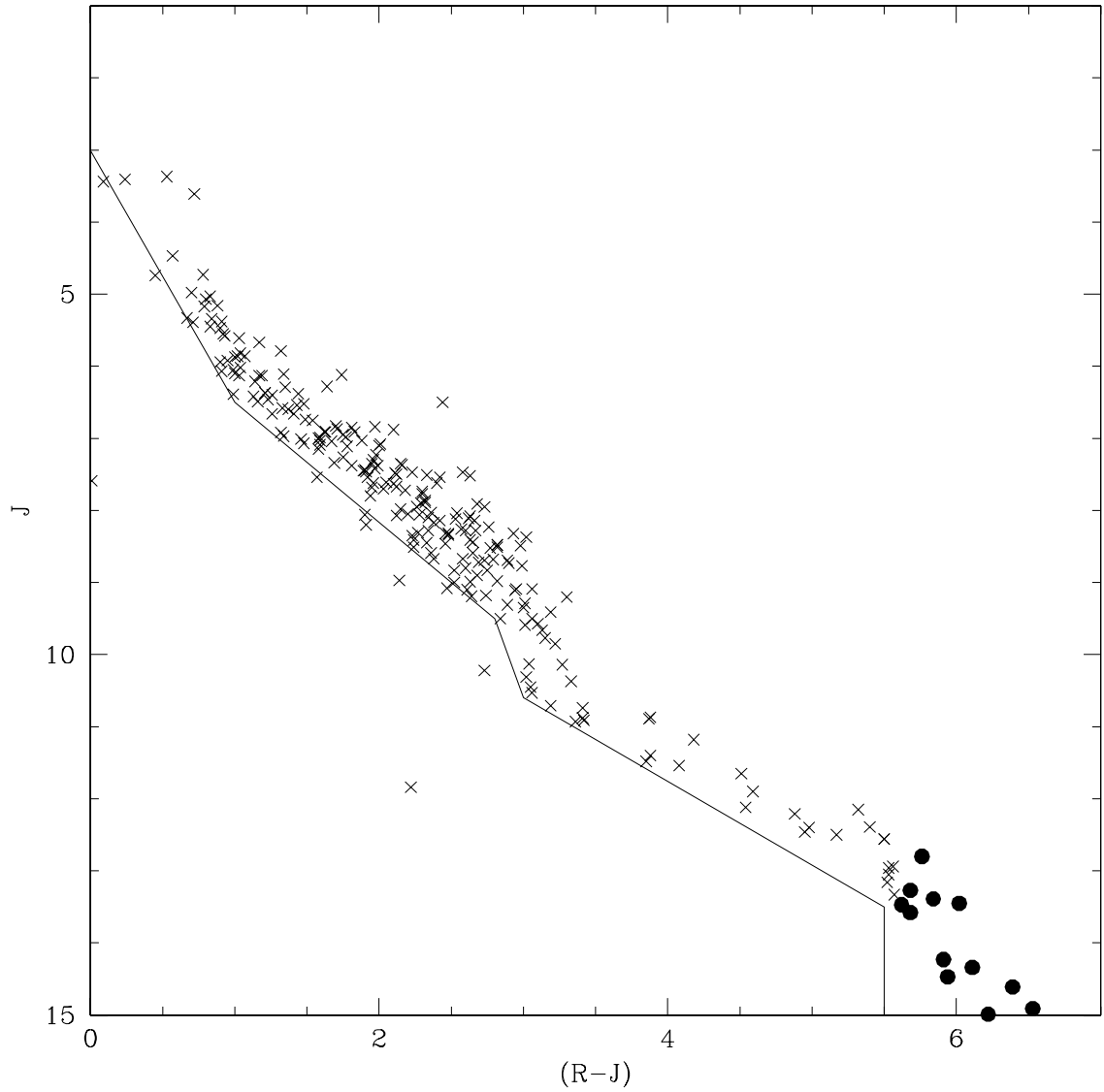}

\caption{Color-magnitude diagram for GKM dwarfs (crosses) and L dwarfs
(solid circles) with known parallaxes and shifted to 20 pc. The solid line
shows the cuts.  Targets are selected if they lie above the
cut.}\label{fig:rj}
\end{figure}

\begin{description}
\item[Cataloged Cloud and Dense Regions:] Objects which are
associated with star-formation regions (e.g. Orion, Lupus, etc.)
were eliminated from the outset. Rough positions and dimensions
for those reddening regions were obtained from \citet{Dame} and
\citet{CDN}.  However, since the high density of sources
persisted on the fringes of those regions, we enlarged the areas
excluded from our sample.  The positions of those regions are
listed in Table \ref{tab:red1}.  In addition, the dense stellar
associations of the LMC, SMC, 47 Tuc, M31, and M33 were also
excluded.  A total of $\sim$1.65 million sources were eliminated
based on these, reducing the sample to $\sim$9.65 million
targets.  These cuts are taken into account in computing the
total areal coverage of the final 2MU2 sample.

\item[\emph{J/(J$-$K$_S$)}:] Our goal is to identify nearby
objects of spectral types M7 and later.  This is accomplished
with two cuts in $J$/($J-K_S$). The left panel of Figure
\ref{fig:fig1} is the color-magnitude diagram for the tail end of
the main sequence as it would appear if all of the objects were
at 20 pc ($M_J + 1.51$).  A color-magnitude diagram for a typical
$1\degr$ region of the 2MASS database is shown in the right
panel. We selected objects that meet both of the following
criteria:
\begin{equation}
\begin{array}{c}
(J-K_S) > 1.0 \\
J\leq1.5(J-K_S)+10.5.
\end{array}
\end{equation}
The ($J-K_S)>1$ criterion eliminates most objects earlier than M7.
Objects that are farther away than 20 pc should fall below the
line formed by the second criterion while nearby objects lie above
it.  These criteria have the biggest effect on narrowing our
sample by cutting $\sim$9.41 million sources, leaving
$\sim$236,500.

One calibrating object (2M 1632, L8) lies just below our line in the
$JK_S$ plane and just above the scatter of the main sequence easily
visible in the right panel of Figure \ref{fig:fig1}.  In our effort to
balance sample completeness with manageable size, we made the $JK_S$ cut
such that numerous faint main-sequence stars would be excluded rather than
accommodate this blue-ish L8 dwarf.

These criteria were not intended to select for T dwarfs however
some early-type examples may fall within our sample (see 2M 0423
in \S \ \ref{sec:near}).

\item[\emph{(H$-$K$_S$)/(J$-$H)}:] To further refine the spectral
type distribution of the sample, color cuts were applied in the
$JHK_S$ plane.  Figure \ref{fig:fig2} shows the color-color
diagram for the same objects as in Figure \ref{fig:fig1} and our
selection criteria:
\begin{equation}
\begin{array}{cl}
(J-H) \le 0.8 & \mbox{for} \; 0.30 \le (H-K) \le 0.35 \\
\left.
\begin{array}{l}
(J-H) \leq 1.75(H-K_S)  +  0.1875 \\
(J-H) \geq 1.75(H-K_S)  -  0.4750
\end{array}
\right\} & \mbox{for} \; 0.35 < (H-K) \le 1.20.
\end{array}
\end{equation}
In addition, the $(J-K_S)>1$ criterion translates to
\begin{displaymath}
(J-H) > 1-(H-K)
\end{displaymath}
in $JHK_S$ color-color space.  These criteria isolate late-M and
L dwarfs and exclude the central region of the giant sequence.
This reduced the sample by $\sim$228,000 objects, leaving 8531
targets.

Figure \ref{fig:fig2} plots these objects from \citet{K99,K00}
whose uncertainties are less than 0.1 magnitudes.  About 15\% of
all of the L dwarfs listed in \citet{K99,K00} scatter outside of
our $JHK_S$ cuts.  The objects that fall outside of our criteria
are among the faintest known L dwarfs with large uncertainties in
their colors (between 0.1 and 0.3 magnitudes).  Extending the
sample to include those regions is not feasible since it would
introduce an unmanageable number of unwanted interlopers.

\item[\emph{J/(R$-$J)}:]Included in the 2MASS data products is
$BR$ optical photometry for point sources with a counterpart in
the Tycho or USNO catalogs within $5\arcsec$ of the 2MASS
position. For objects where an $R$ magnitude is available, we have
applied a cut in $J/(R-J$) to restrict the sample to objects with
optical/NIR colors that are consistent with nearby late-type
dwarfs.  Figure \ref{fig:rj} shows the color-magnitude diagram for
cool stars with known parallaxes where the apparent magnitudes
have been shifted to as they would appear if at a distance of 20
pc ($M_J+1.51$).  Targets with an available $R$ magnitude
($\sim40\%$) are required to meet the following criteria:
\begin{eqnarray}
J \leq  \left\{
\begin{array}{ll}
3.50(R - J) + 3.00, &  \mbox{for} \; (R - J) \leq \ 1.0, \\
1.67(R - J) + 4.83, &  \mbox{for} \; 1.0 < (R - J) \leq 2.8, \\
5.50(R - J) - 5.90, &  \mbox{for} \; 2.8 < (R - J) \leq 3.0, \\
1.16(R - J) + 7.12, &  \mbox{for} \; 3.0 < (R - J) \leq 5.5. \\
\end{array}
\right.
\end{eqnarray}
This cut 5337 objects, leaving 3194 targets.

\item[Giants:]  \citet{bb} have shown that some types of giants
have colors that meet our $JHK_S$ color criteria and as a result,
many M giants persisted in the sample.  Figure \ref{fig:fig2}
shows data for giants in $JHK_S$ color space.  Noting that giants
tend to have bluer ($H-K_S$) colors and very bright magnitudes
($J<10$), we are able to get rid of a significant portion of the
giants by eliminating objects that meet both of the following
criteria. (Note that the previous items have been criteria for
inclusion while the following criteria are for exclusion.)
\begin{equation}
\begin{array}{c}
J < 10 \\
(J-H) > 2(H-K_S), \; \mbox{for} \; 0.375 < (H-K_S) < 0.470.
\end{array}
\end{equation}
This cut 698 objects, leaving 2496 targets.  The second criterion
is shown in Figure \ref{fig:fig2}.

\item[Flags:]  The 2MASS catalog also provided flags for possible
photometric confusion and solar system objects (cc\_flg=000 and
mp\_flg=0).  The confusion flag indicates when more than one
object is merged and the individual photometry could not be
resolved.  The solar system flag indicates the point source is
associated with a minor planet.  Three hundred ten objects were
cut based on these flags, leaving 2186.

\item[Uncataloged Reddening Regions:]  Even though we cut objects
at low galactic latitudes and near star-forming regions, there
were still several large areas of high density near the plane.
These regions are listed in Table \ref{tab:red2} and are likely
to be reddened sources associated with small/uncataloged
molecular clouds, so we have eliminated them.  This removed 514,
leaving a sample size of 1672 targets.

\end{description}

\subsection{Refining the Faint Portion of the 2MU2 Sample}

\begin{figure}
\epsscale{1.0} \plotone{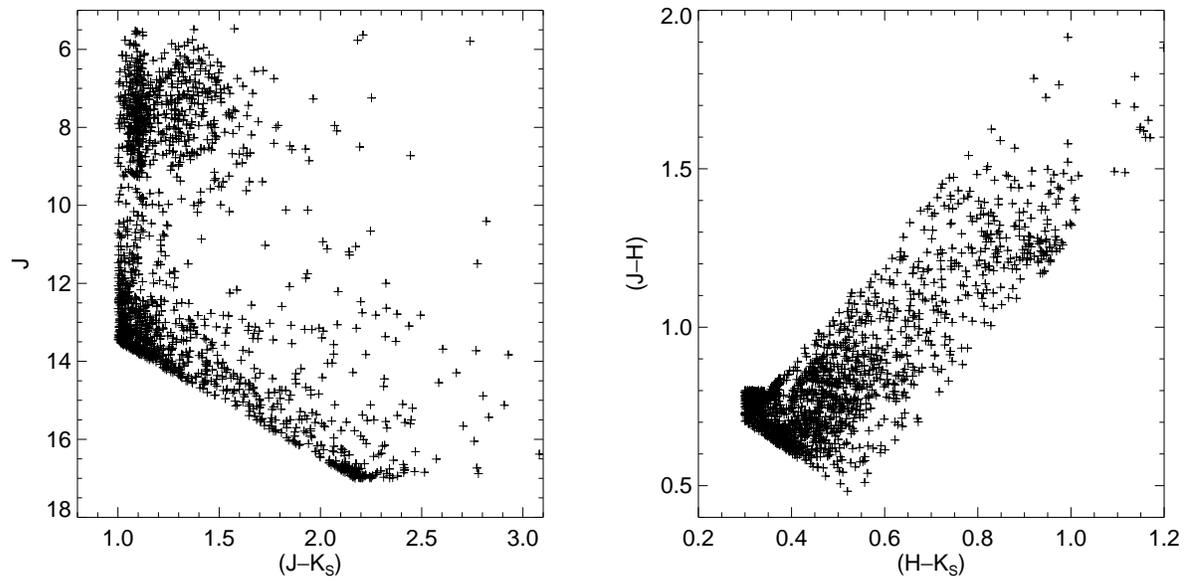}

\caption{Color-magnitude and color-color diagrams for the targets that
survived the cuts described in \S\ \ref{sec:cuts}.  Objects fainter than
$J=9$ are candidates for spectroscopic follow-up.  The population of
objects with $J\le9$ are discussed in \S\
\ref{sec:bright}.}\label{fig:fig3}
\end{figure}

The color-color and color-magnitude diagrams of the 1672
surviving candidates are shown in Figure \ref{fig:fig3}.  At this
point, the 588 objects with $J\le9$ were set aside for separate
classification and are discussed in \S\ \ref{sec:bright}. The
1084 objects with $J>9$ were examined in more detail and 447
objects were eliminated based on the following considerations
which are summarized in Table \ref{tab:morecuts}:

\begin{description}

\item[\emph{(F$-$J)/(J$-$K$_S$)}:] Additional optical photometry
was obtained from the Guide Star Catalog 2.2 \citep[GSC]{gsc}.
($F-J$) and ($J-K_S$) colors were used to eliminate objects that
are too blue at optical wavelengths to be ultracool dwarfs.
Targets were required to meet the following criteria:

\begin{eqnarray}
(F-J) > \left\{
\begin{array}{l}
4.0 \\
1.67(J-K) + 2.33.\\
\end{array}
\right.
\end{eqnarray}

This eliminated 137 objects.  Recently, \citet{LK03}, identify one
of these eliminated objects as a nearby L dwarf. As suggested by
those authors, crowding in the field likely resulted in a mismatch
between the 2MASS source and the GSC source and thus an aberrant
$(F-J)$ color.  All of these 137 objects will be examined by eye
to correct for other possible mismatches and will be discussed in
a future paper.

\item[Visual Inspection:] Two hundred and eleven targets were
eliminated via visual inspection of POSS plates and 2MASS images.
The images revealed some sources to be artifacts of the 2MASS
data, associated with large galaxies or globular clusters, or
clearly non-red objects.

\item[SIMBAD:] Twenty four objects were eliminated because they are listed
in SIMBAD as known carbon stars, pre-main sequence objects, or quasars.

\item[Clouds:] Both SIMBAD and visual inspection enabled us to
identify 41 objects associated with molecular clouds or reddening
regions.  These were eliminated.

\item[Bland:] Figure \ref{fig:fig3} shows the color-magnitude and
color-color diagrams for the 588 bright objects as well as the surviving
faint targets.  As can be seen in the left panel there is a high density
of faint objects with colors close to $(J-K_S)=1$.  The high density of
mid-M dwarfs just bluer than $(J-K_S)=1$ and the increased photometric
uncertainties at fainter magnitudes leads to more scatter into our sample.
After many observations of objects in this color space, we were able to
confidently eliminate 34 objects with $J-K_S<1.1$ and $J>13$ as distant
mid-M dwarfs.
\end{description}

These cuts reduced the 2MU2 sample of $J>9$ objects to 637
ultracool candidates. Table \ref{tab:known} lists the 112
ultracool targets with existing data (predominately from
\citet{K99, K00, NN}).  The remaining 525 require spectroscopic
follow-up observations---data for 307 are presented here.

\subsection{The Brightest Candidates}
\label{sec:bright}

Five hundred and eighty-eight of the ultracool dwarf candidates
have apparent magnitudes $J\le9$. Based on both the
color-magnitude distribution, and the fact that most lie
relatively close to the Galactic plane, we discuss these sources
separately.

Cross-referencing against SIMBAD, using the 2MASS positions,
leads to positive identifications for 386 sources, as follows:

\begin{itemize}

\item One hundred seventeen sources are within 10--15\arcsec\ of
a source from the IRAS catalog. These are likely to be dusty
giants, asymptotic giant branch (AGB) stars, supergiants or young
protostars. Again, most lie close to the Plane.

\item One hundred thirty eight sources lie within 2--3\arcsec\ of
stars cataloged in the Henry Draper, Bonner Durchmusterung, Cape
Durchmusterung, Cape Photographic Durchmusterung, Guide Star
Catalog, or PPM catalog. None of these stars have significant
proper motions, and the near-infrared and optical/infrared colors
are consistent with red giant or AGB stars.

\item Thirteen stars are classified in SIMBAD as carbon stars.
Again, all lie on the AGB.

\item Eighty stars are identified as Miras and sixteen stars are
classed as semi-regular variables; both datasets are
predominantly M-type AGB long-period variables. These stars show
a strong concentration towards the Plane and the Bulge.

\item Eighteen sources are identified with a variety of late-type
stars, including red giant variables, symbiotic stars, pre-main
sequence variables and a symbiotic star.

\item Finally, four sources are matched to known M-dwarf proper
motion stars. We discuss these further below.

\end{itemize}
Near-IR color-magnitude diagrams of these sources are shown in the
Appendix.

Two hundred and two bright ultracool candidates have no previous
identification listed in the SIMBAD database. However, all save
two of those sources have optical counterparts (positional offset
of less than 5\arcsec\ from the USNOA-1.0 catalog listed in the
2MASS database. Since the USNO catalog is based on scans of the
POSS I $O$/$E$ plates in the North (average epoch $\sim$1953) and
the SERC/UKST $J$ and ESO $R$ plates in the South (average epoch
$\sim1979$), this indicates that these sources have proper motions
of less than 0\farcs1 yr$^{-1}$ and 0\farcs2 yr$^{-1}$,
respectively---indeed, in every case the measured offsets indicate
negligible relative motion (USNO/2MASS). Both sources which lack
cataloged USNOA-1.0 counterparts are clearly visible at the 2MASS
position on second-epoch UKST IIIaF plates taken in the mid-1980s,
also indicating low proper motions.

\begin{figure}
\epsscale{1.0} \plotone{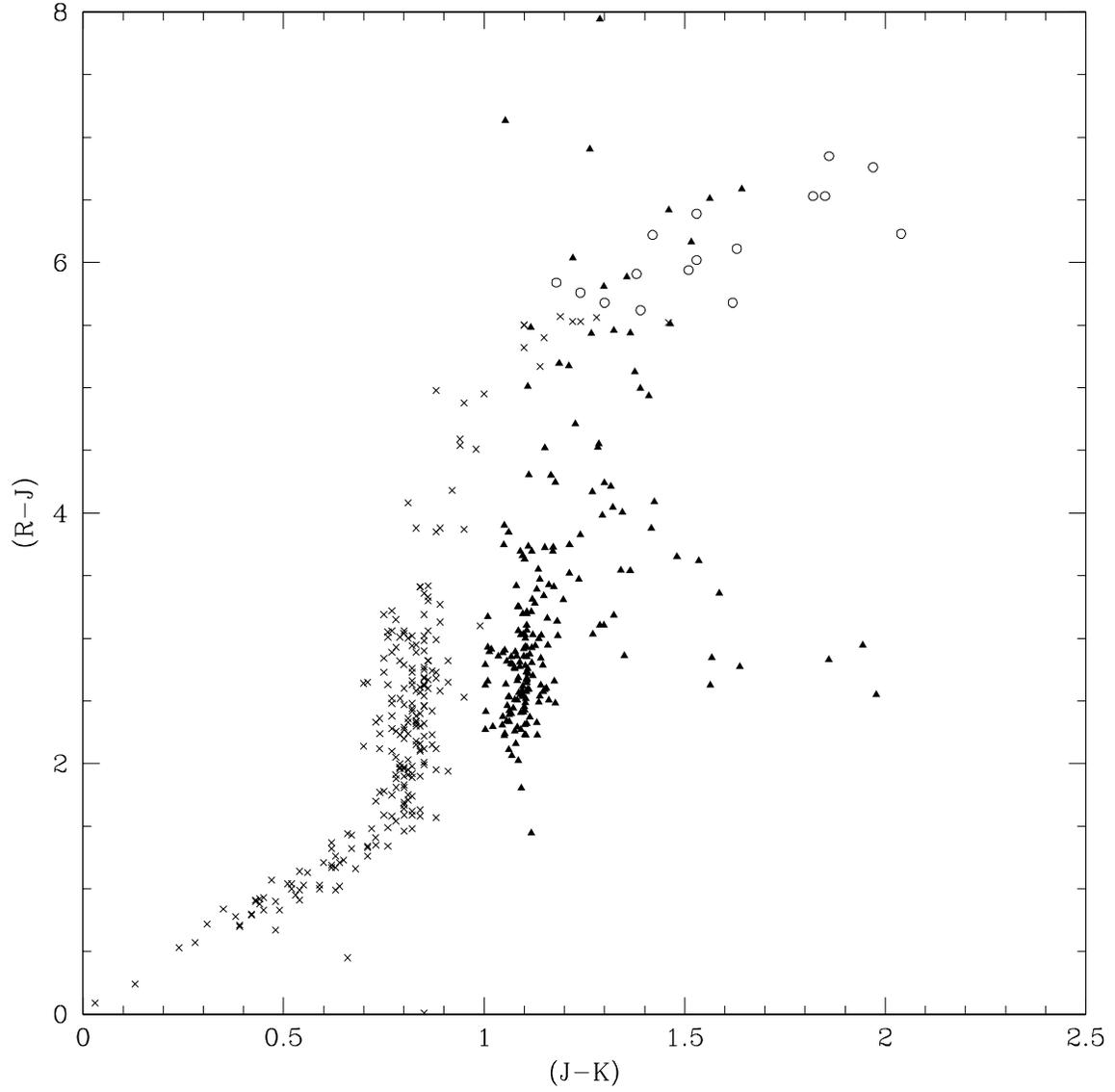}

\caption{The ($R-J$)/($J-K_S$) distribution for main-sequence FGKM stars
(crosses), L dwarfs (open circles) and bright ultracool candidates with no
counterpart listed by SIMBAD (solid triangles).} \label{fig:bright2}
\end{figure}

The presence of non-moving optical counterparts  strongly suggests
that all of these uncataloged sources are likely to be distant
reddened stars or red giants, rather than nearby dwarfs.
Confirmation of this hypothesis comes from the ($R-J$)/($J-K_S$)
two-color diagram (Figure \ref{fig:bright2}), which shows that the
overwhelming majority have significantly bluer ($R-J$) colors than
expected for late-M or L dwarfs. Given the low Galactic latitude
of most sources, the potential exists for mismatches. However, as
discussed in more detail in the Appendix, spectroscopy of the
relatively small number of sources with near-infrared colors
consistent with late-M/L dwarfs shows that none are late-type
dwarfs.

In summary, only four sources among the 588 ultracool candidates
with $J\le9$ are likely to be genuine nearby late-type dwarfs.
These stars are as follows:
\begin{description}

\item[G 180-11:]  This object is included in the third Catalog of
Nearby Stars \citep[pCNS3]{pCNS3}. \citet{PMSU} measure a spectral
type of M4.5 and estimate a distance of 13$\pm$3.9 pc.
($M_V=13.11$). \citet{MZB} identify a candidate companion, with
$I=12.6$ and a separation of 1\farcs5 at PA=266$\degr$.  There is
no object at the appropriate location on the POSS I image of this
field, suggesting that the fainter object is associated with G
180-11.  If the companion were a late-type dwarf, then the
absolute magnitude ($M_I \sim 12$) implies a spectral type of
$\sim$M6 and $K_S \sim 10$, $\sim2$ magnitudes fainter than the
primary.  There is no evidence for significant distortion in the
2MASS point-spread function, but that might reflect the large
pixel scale (1\farcs5) and consequent poor sampling.
Alternatively, the hypothesized companion might be a white dwarf,
in which case the inferred absolute magnitude would be consistent
with T$_{eff}\sim 11,000$ K and an expected apparent magnitude of
$J\sim12.5$. Further observations are required to decide between
these hypotheses.

\item[G 139-3:] This is also a star included in the pCNS3. The
spectral type is M4 \citep{PMSU}, and the spectroscopic parallax
gives a distance of 13.8$\pm$4.0 pc ($M_V=12.49$).

\item[BD-01 3925D:]  This object is a wide companion of the K0
dwarf,  BD -1:3925A or HD 192263, ($\pi=50.27\pm1.13$ mas;
\citet{hipparcos}).  BD-01 3925D is not listed in the pCNS3, but
has $M_K=5.61$ for $d=19.9$ pc, consistent with a spectral type
of M0/M1.

\item[EZ Aqr:] This object is also known as Gl 866, the
well-known triple system \citep{Delfosse99}. This system is
included in pCNS3, has a spectral type of M5.5, and a distance of
3.4$\pm$0.03 pc.

\end{description}

With the possible exception of Gl 866ABC, these dwarfs have a
relatively unusual location on the ($J-H$)/($H-K_S$) diagram.  In
the 2MASS All-Sky Release (where the photometry for bright stars
is improved over the Second Release), the revised photometry for G
180-11 and G139-3 puts them blueward of our $J-K=1$ cut but the
remaining two objects still have unusually red colors.  Since M
dwarfs with spectral types earlier than M5 would be expected to
have ($J-K_S)< 1.0$, further observations of these objects are
desirable.  Data for these stars are listed in Table
\ref{tab:bright}.

\section{Observations}
\label{sec:obs} Follow-up far-red optical spectroscopy has been
obtained for 298 potential nearby dwarfs during five separate
observing runs at NOAO facilities\footnote{Spectra are available
upon request from \email{kelle@sas.upenn.edu}.}.  (Data for 9
objects were obtained from other sources.) Tables
\ref{tab:latenear}--\ref{tab:carbon} list the positions (as a
2MASS name), photometry, observation date, and derived data for
all of the observed targets\footnote{The 2MASS designation is
2MASSI Jhhmmss[.]s$\pm$ddmmss.  In addition, we note that the
astrometry and photometry for all objects is likely to be
different from those listed in the 2MASS All-Sky Release.}.

The first run was at the Kitt Peak 2.1 m with the GoldCam
spectrograph on 2000 September 29 through October 2. The majority
of the run was dedicated to follow-up of proper-motion selected
candidates (Paper III).  We used a 1$\farcs$3 slit and a 400 line
mm$^{-1}$ grating blazed at 8000 \mbox{\AA} to give a resolution
of 5.1 \mbox{\AA} (2.8 pixels) covering the wavelength range
5500--9300 \mbox{\AA}.  Higher orders were blocked using a OG550
filter.  Weather was good with 1\arcsec--1$\farcs$5 seeing.  These
data were not flat-fielded due to problems with fringing (see
Paper III for a detailed discussion).

Observations were obtained on the Kitt Peak 4 m on 2001 July
13--23 with the RC Spectrograph to cover 6000--10000 \mbox{\AA}
in first order.  An OG530 filter was used to block second-order
light.  We achieved a resolution of 5.6 \mbox{\AA} (2 pixels)
with a 316 line mm$^{-1}$ grating blazed at 7500 \mbox{\AA} and a
1$\farcs$5 slit.  The run had fair weather with partly cloudy
conditions and a wide range of seeing.  We were able to confirm
many carbon stars and giants as well as obtain data on brighter
targets.

The same instrumental setup was used on 2002 January 21--24
except an OG550 filter was used to block higher orders.  Good
weather (seeing 0$\farcs$9--1\farcs2) on the first two nights and
the fourth night permitted the use of a 1$\arcsec$ slit resulting
in a resolution of 4.7 \mbox{\AA} (1.7 pixels).  The seeing on
the third night ranged from 1$\farcs$5--2$\arcsec$ and a
1$\farcs$5 slit was used to obtain a resolution of 5.2 \mbox{\AA}
(1.9 pixels).

Coincident observing (2002 January 22--25) with the Blanco 4 m on
Cerro Tololo also had good weather with mostly
1$\arcsec$--2$\arcsec$ seeing.  A Loral 3K CCD and the RC
spectrograph with a 315 line mm$^{-1}$ grating blazed at 7500
\mbox{\AA} and a 1$\arcsec$ slit was used to cover the range
5500--10000 \mbox{\AA} with a resolution of 5.5 \mbox{\AA} (2.8
pixels).  An OG515 filter was used to block higher orders.

Immediately following, data were obtained with the CTIO 1.5 m on
2002 January 26--30 with a Loral 1K CCD and RC Spectrograph to
cover 6000--8600 \mbox{\AA}. Resolution of 6.5 \mbox{\AA} (3
pixels) was obtained with a 400 line mm$^{-1}$ grating blazed at
8000 \mbox{\AA} and a 1\farcs5 slit. The seeing ranged from
0\farcs7--2\arcsec.

All the spectra were flat-fielded, corrected for bad pixels,
extracted, and wavelength- and flux-calibrated using the standard
IRAF packages CCDPROC and DOSLIT.  (Data from 2000 September were
not flat-fielded.)  Wavelength calibration was determined using
HeNeAr arcs taken nightly.  The flux-calibration was done using HD
19445, HD 84937, Ross 640, G 191-B2b, L 1363-3, and Feige 34
\citep{Oke,Hamuy}.

\section{Results}
\label{sec:results} 
\begin{figure}
\epsscale{1.0} \plotone{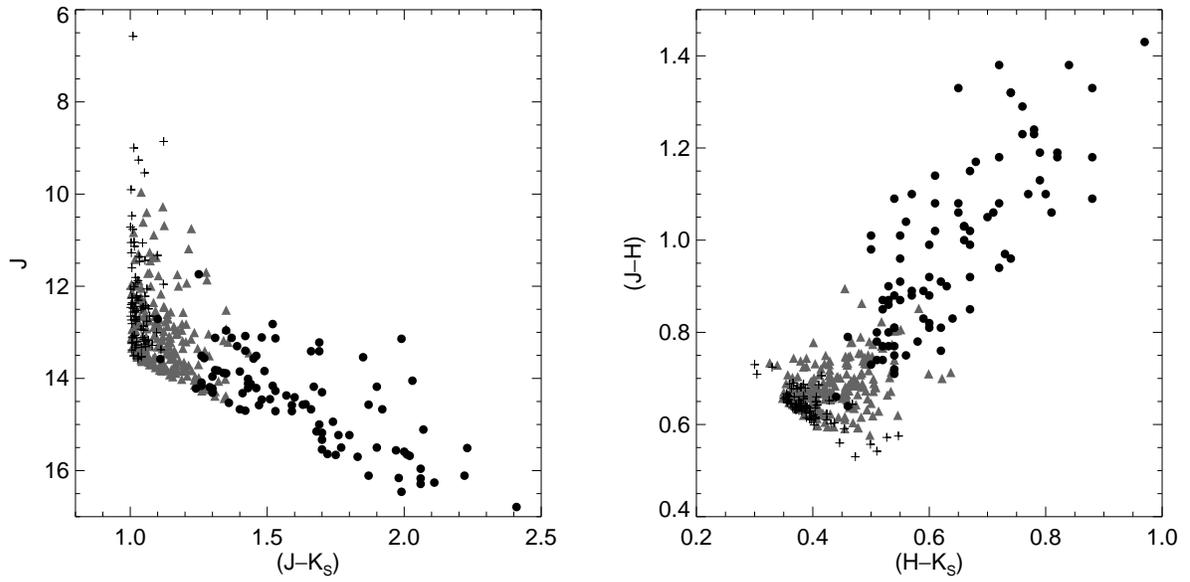}

\caption{Color-magnitude and color-color diagrams for the all of the cool
dwarfs present in this Paper. Crosses are early and mid-M dwarfs
(M0--M6.5), filled triangles are late-type M dwarfs (M7--M9.5), and filled
circles are L dwarfs.  The two blue L dwarf outliers are discussed in \S\
\ref{sec:blue}}\label{fig:done}
\end{figure}

We have derived spectral types, absolute
magnitudes, and distances for all of the observed dwarfs. Absolute
magnitudes are estimated based on a spectral type/$M_J$ relation;
and distance estimates are made using the derived $M_J$ and 2MASS
$J$-band photometry.  Table \ref{tab:latenear} lists the data for
forty-seven newly discovered objects that lie within 20 pc and
have types M7 and later. These objects are additions to the core
sample that is used for our luminosity and mass function analysis
(see \S \ref{sec:lf}. Data for 139 ultracool dwarfs that lie
outside of 20 pc are listed in Table \ref{tab:latefar}. Data for
ten M dwarfs earlier than M7 with a distance estimates less than
20 pc (or within 1$\sigma$) are listed in Table
\ref{tab:earlynear} while fifty-three more distant objects are
listed in Table \ref{tab:earlyfar}.  Five objects that have
spectral features indicative of youth are listed in Table
\ref{tab:young}, fifty-four giants and carbon stars/dwarfs are
discussed below and listed in Tables \ref{tab:giants} and
\ref{tab:carbon}.  Three objects (\objectname{2MASSI
J0150116+152357}, \objectname{ 2MASSI J0447575$-$055324}, and
\objectname{2MASSI J0442007$-$135623}) are reddened, distant,
early-type stars.   The color-color and color-magnitude diagrams
for the 365 dwarfs present in the portion of the 2MU2 sample
presented here are shown in Figure \ref{fig:done}.

\subsection{Spectral Types}
We found that using our spectral indices (listed in Table 4 of
Paper III) as a predictor of spectral type was unreliable for cool
dwarfs. This is especially true for M6--M8 dwarfs since the TiO 5
relation turns around near M7.5 (see Paper III, Figure 3, left
panel), leading to highly ambiguous classifications.

As a result, all spectral types for M dwarfs in this Paper are
determined via visual comparison with standard star spectra taken
during the course of our program.  Program objects are typed by
being normalized and plotted between spectra from a grid of eight
standard M1--M9 dwarfs \citep{KHM91}.  L dwarf types are
determined via comparison with nine standard, integer type L0--L8
publicly available LRIS spectra taken as part of the 2MASS Rare
Objects project \citep{K99,K01}.  In addition, integer types are
favored over half-integer types.  The resulting uncertainty for
all types is $\pm$0.5 subtypes, except where noted by a ``?''
where low signal-to-noise increases the uncertainty to 1 or 2
types.

\subsubsection{Giants and Carbon Stars/Dwarfs}

In the course of spectroscopic follow-up of potential nearby
dwarfs, many targets turned out to be distant giants or carbon
stars.  In Table \ref{tab:giants} we list the photometry,
astrometry, and rough spectral types ($\pm$1 type) for the
observed giants.  The spectral types were estimated by
side-by-side comparison with observed spectral standards from
\citet{giants} and spectra kindly provided by J.D. Kirkpatrick
from \citet{KHI}.  In Table \ref{tab:carbon} we list the
photometry and astrometry for the carbon stars.  Two of these are
carbon dwarfs with detectable proper motion between the first
epoch sky survey plates and the 2MASS images.  These will be
discussed in detail in a future paper (P.J. Lowrance et al., in
preparation).

\subsection{Absolute Magnitudes and Distances}
\label{sec:cal} $M_J$ and distance estimates for all of the
dwarfs in our sample are listed in Tables
\ref{tab:known}--\ref{tab:earlyfar}. Absolute magnitudes for
spectral types M2--M5.5 were derived using TiO 5, CaH 2, and CaOH
as described in Paper III (Figure 4). For later types, however,
the index relations are double valued.  As a result, we have used
spectral type as a predictor of $M_J$.

\begin{figure}
\epsscale{1.0} \plotone{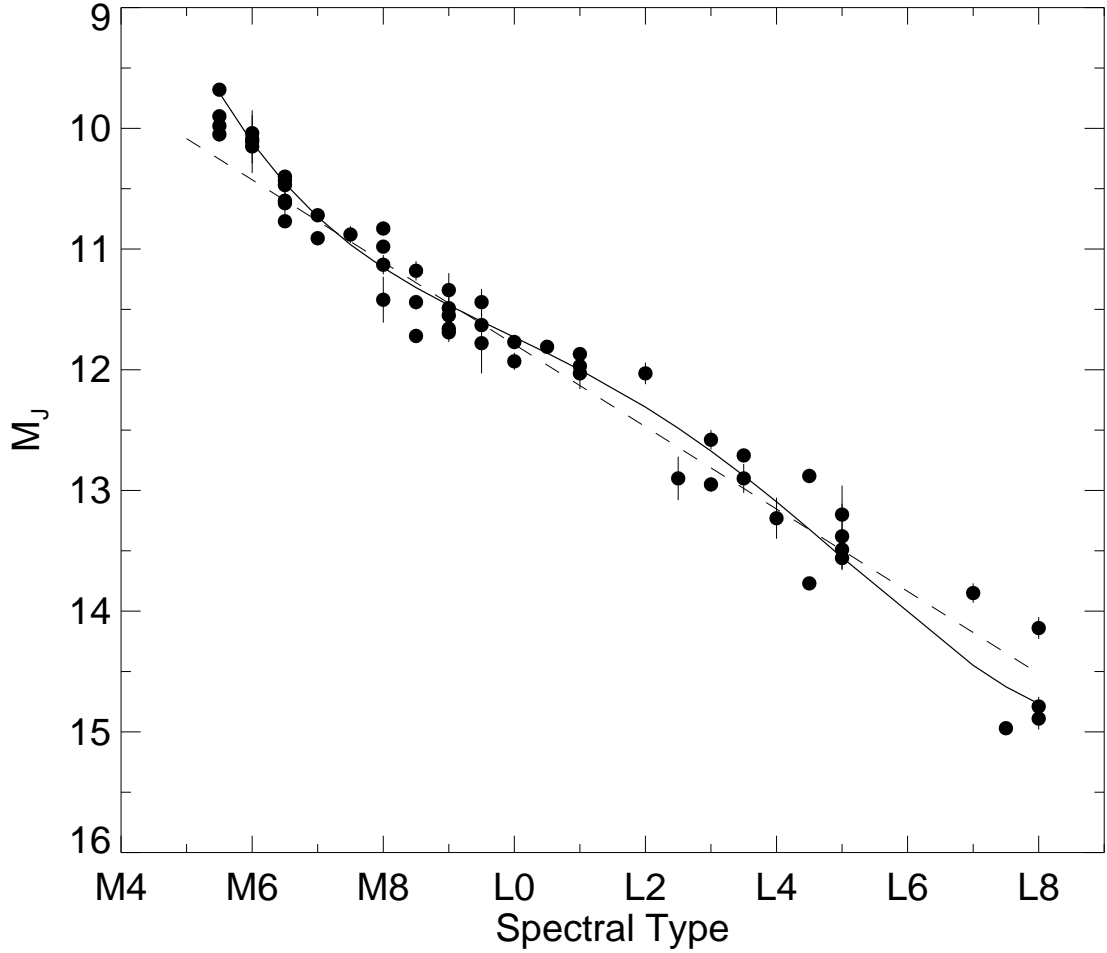}

\caption{$M_J$/spectral type calibration.  The solid line is our fourth
order fit to the data, while the dashed line is the linear fit found by
\citet{Dahn} to a similar dataset.  Data plotted are listed in Table
\ref{tab:mjcal}} \label{fig:mjcal}
\end{figure}

\citet{Dahn} have shown that $M_J$ is well correlated with
spectral type.  We have rederived that relation, supplementing
data kindly provided by H. Harris with several early objects from
PMSU and the 8 pc samples. The data used to make the relation are
listed in Table \ref{tab:mjcal} and are plotted in Figure
\ref{fig:mjcal}. A fourth order polynomial was found to best fit
the data, and is valid for types M6--L8:
\begin{equation}
M_J = -4.410 + 5.043(ST)
-0.6193(ST)^2+0.03453(ST)^3-0.0006892(ST)^4
\end{equation}
where ST=0, 10, 18 for spectral types M0, L0, L8, respectively.
The uncertainty in spectral type dominates the uncertainty in the
estimated $M_J$.

Distances are derived for program objects using the estimated
$M_J$ and 2MASS $J$ photometry.  The uncertainties in the derived
distances are also dominated by the uncertainty in spectral type.
(The uncertainties in the 2MASS $J$ photometry are typically only
0.02--0.03 magnitudes and, as stated above, the uncertainty in
$M_J$ is mostly due to the uncertainty in spectral type.)

\section{Interesting Individual Objects}
\label{sec:obj}  
\begin{figure}
\epsscale{1.0} \plotone{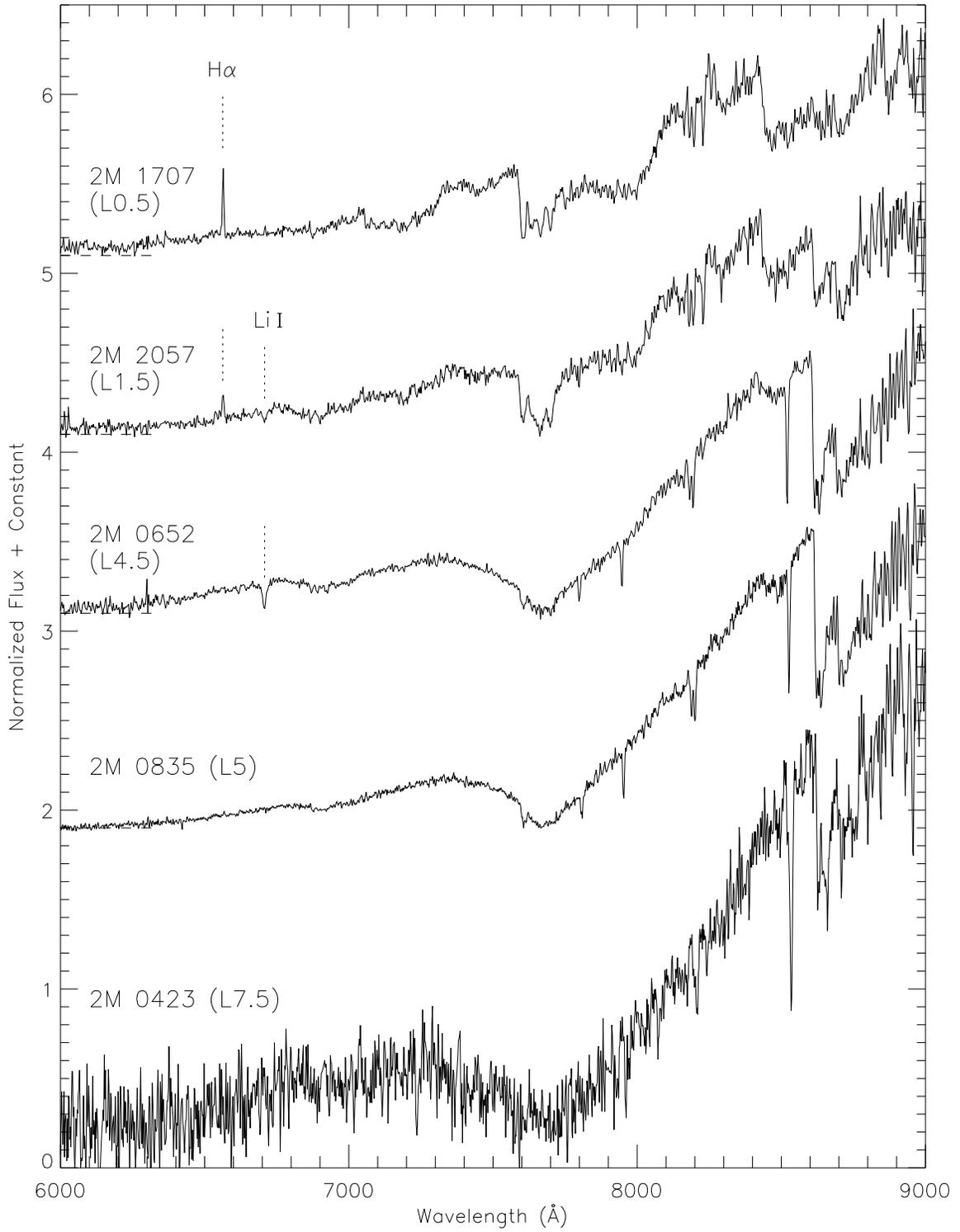}

\caption{Spectra for some of the interesting objects.  Each object is
discussed in \S\ \ref{sec:obj}. The bottom spectrum is not offset and the
zero point for each offset spectrum is shown by a dashed
line.}\label{fig:interesting}
\end{figure}

\begin{figure}
\epsscale{1.0} \plotone{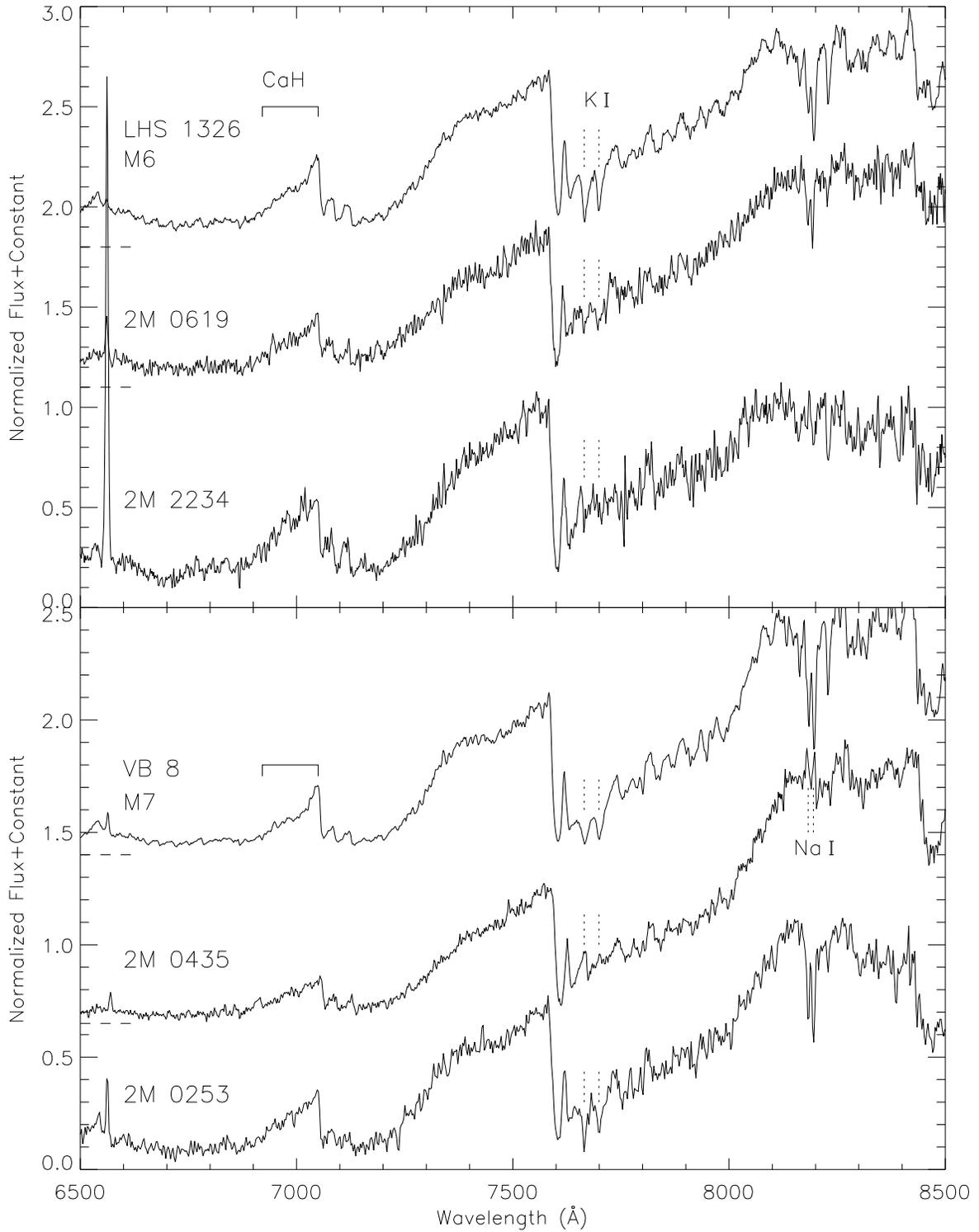}

\caption{Spectra for candidate young objects and the M6 and M7 spectral
standards LHS 1326 and VB 8.  The spectral features of each object are
discussed in \S \ref{sec:young}.  The bottom spectrum is not offset and
the zero point for each offset spectrum is shown by a dashed
line.}\label{fig:young}
\end{figure}

\begin{figure}
\epsscale{1.0} \plotone{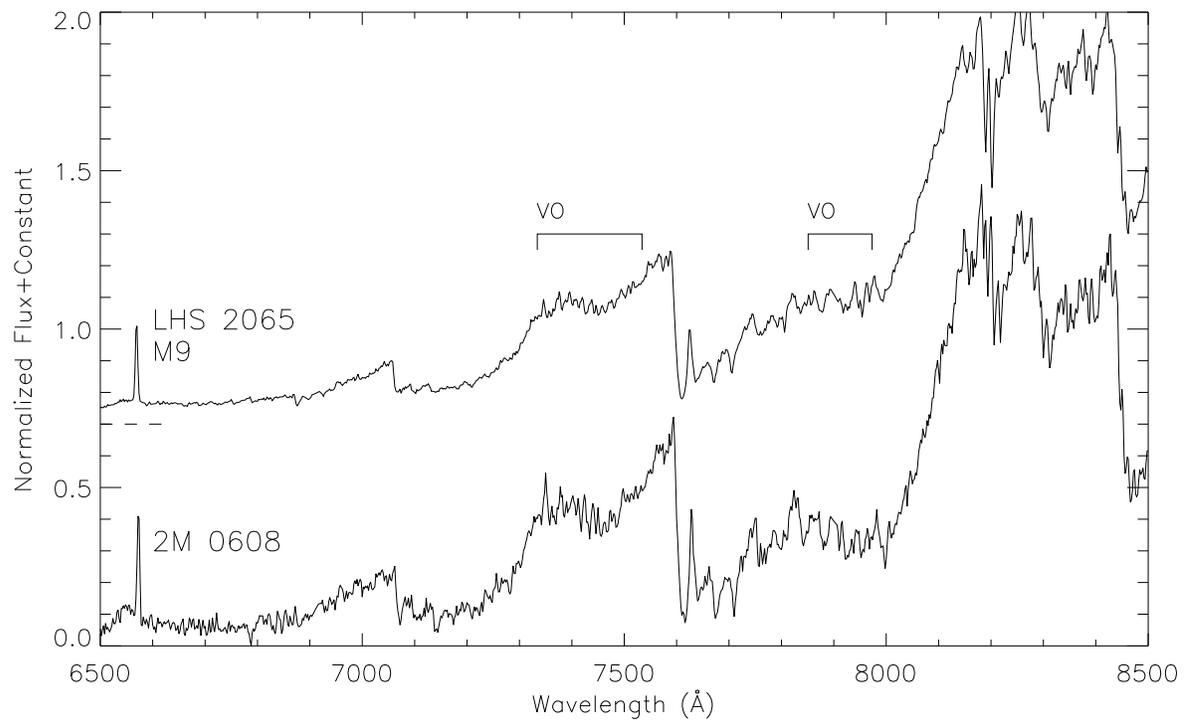}

\caption{Spectra of 2M 0608 and the M9 standard LHS 2065.  This object is
discussed in \S \ref{sec:young}.  The bottom spectrum is not offset and
the zero point of the offset spectrum is shown by a dashed
line.}\label{fig:young2}
\end{figure}

Spectra for many of these objects are shown in
Figures \ref{fig:interesting} -- \ref{fig:young2}.

\subsection{L Dwarfs within 10 pc}
\label{sec:near}
\begin{description}

\item[2M 0423:] L7.5 at 9.2 pc.  This object is typed as T0 by
\citet{Geballe} based on the strength of water absorption and the
detection of methane absorption bands in the 1--2.5 \micron\
region.  In addition, \citet{SDSS2} type this object as L5: based
on low-resolution optical spectra.  In general, spectral types
based on optical and near infrared spectra agree, but there are a
few cases where they do not.  This issue is discussed in
\citet{B03} and will be addressed in detail in a future paper
(J.D. Kirkpatrick et al., in preparation).

\item[2M 0835:] L5 at 8.3 pc.  This L dwarf is among the brightest
of its type, with $J=13$.  As such, it is a good candidate for
high resolution observations to check for a possible companion.  A
trigonometric parallax measurement would also help to confirm this
as a single L dwarf within 10 pc or a multiple system at a
slightly larger distance.
\end{description}

\subsection{Brown Dwarfs}
\label{sec:li} Two very nearby L dwarfs in our sample have
\ion{Li}{1} absorption at 6708 \mbox{\AA} confirming their brown
dwarf status.

\begin{description}

\item[2M 0652:] L4.5 at 11.1 pc with strong lithium absorption
feature.

\item[2M 2057:] L1.5 at 15.7 pc.  This spectrum displays both
lithium absorption and H$\alpha$ emission.  Kelu-1 and the
previously mentioned 2M 0423 also both have these features
(\citet{Kelu1} and J.D. Kirkpatrick et al., in preparation).

\end{description}

\subsection{Active Objects}
\label{sec:active} We observed several objects that had unusually
strong H$\alpha$ emission---we list the equivalent width (EW) of
the line below.  The activity level of the entire sample will be
discussed in a future paper (J. Liebert et al., in preparation).
\begin{description}

\item[LP 423- 31/2M 0752:] This M7 (at 10.5 pc) was observed three
times during our program: twice in 2002 January and once during a
run dedicated to NLTT follow-up in 2001 November.  On all
occasions, the emission strength was very strong with the EW
ranging from 33--45 \mbox{\AA}.

\item[2M 1707:] Emission in L dwarfs is not common---only two
presented here display it.  This L0.5 (at 26 pc) has significant
H$\alpha$ with EW=35 \mbox{\AA} and this object should be
monitored to check whether we happened to observe during a period
of unusually high activity.  A spectrum of this object is shown
in Figure \ref{fig:interesting}.

\item[2M 2057:] This brown dwarf with H$\alpha$ emission is
discussed in \S\ \ref{sec:li} above.

\item[2M 2351:] This M7 at 35 pc displayed a strong H$\alpha$
emission (EW=51 \mbox{\AA}) in 2001 July and, unfortunately, no
further observations were taken.

\end{description}

\subsection{Young Objects}
\label{sec:young}

We have found five objects which have spectral features similar to
young ($\sim 10$ Myr) objects found in the TW Hya association
\citep{Webb}.  However, they do not appear to be associated with
any known star forming regions.  These are listed in Table
\ref{tab:young} and shown in Figures
\ref{fig:young}--\ref{fig:young2} in between standard dwarf
spectra.  All show weak CaH absorption (6750--7050 \mbox{\AA}) and
most show weak \ion{K}{1} doublet absorption (7665 and 7699
\mbox{\AA}) compared to similar solar age dwarfs.  Both of these
features are indicative of low gravity and therefore youth.

\begin{description}

\item[2M 0253:]  From the DSS and 2MASS images, we find that this
object has moved approximately 6$\arcsec$ over 43 years. This
further supports our hypothesis that this is a young nearby dwarf.

\item[2M 0435:] In addition to weak CaH absorption, this object displays
weak emission in H$\alpha$, \ion{K}{1} and \ion{Na}{1}.  It is about one
magnitude brighter than the two TW Hya candidates \citep{twhya} and is
probably within 30 pc. It is in the direction of the nearby star forming
region \objectname{MBM 20}; however, a current estimate of the distance to
this cloud puts it between 112 and 161 pc---significantly more distant
than our estimate for this object \citep{L1642,MBM20}.

\item[2M 0608:] This dwarf also shows enhanced VO absorption
(7334--7534 and 7851--7973 \mbox{\AA}) which is another
characteristic of young late-type objects (J.D Kirkpatrick et al.,
in preparation).  It is also fairly near on the sky to 2M 0619 as
discussed below.

\item[2M 0619:]  This object displays the characteristics of a
young dwarf.  In addition, this object and 2M 0608 are close to
each other on the sky and lie fairly close to the plane.  However,
rough distance estimates of 30 pc for 2M 0608 and 100 pc for 2M
0619 suggest that they are not actually associated.

\item[2M 2234:] This dwarf also displays strong H$\alpha$ emission
consistent with it being a young, active object.

\end{description}

\subsection{Two Blue L Dwarfs}
\label{sec:blue}

There are two L dwarfs which have unusually blue colors and can
easily be seen as outliers in Figure \ref{fig:done}.  One of these
objects, 2MASS J1300425+191235 (L1, 14 pc), was originally
discovered by \citet{NN} and its data are listed in Table
\ref{tab:known}. The other object, 2MASS J172139+334415 (L3, 15
pc) was discovered as a part of this program and its data are
listed in Table \ref{tab:latenear}.

Both of these objects have significant proper motion.  Based on
POSS plates measurements, we find ($\mu_{\alpha},\mu_{\delta})=
-0.7, -1.1 \arcsec \mbox{yr}^{-1}$ for 2M 1300 and
($\mu_{\alpha},\mu_{\delta})=-1.6, 0.6 \arcsec \mbox{yr}^{-1}$ for
2M 1721.  \citet{NN} also pointed out 2M 1300's unusual colors and
high velocity and suggested that it is likely to be old.  The
large proper motion of 2M 1712 supports their claim, however, we
plan to obtain near-infrared spectroscopic observations of these
two objects to find out which spectral features are causing the
anomalous colors.


\subsection{LP 775-31 \& LP 655-48}
Recently, \citet{mccaughrean} pointed out that these two objects
(aka 2MASSI 0435161$-$160657 and 2MASSI 0440232$-$053008) were
mistyped in Paper III and that a retyping places them within 10
pc.  While we agree that these objects are cooler than M6 as we
previously stated, we find them both to be spectral type M7 (not
M7.5 and M8 as in McCaughrean et al.), which puts them at 8.6 and
9.8 pc respectively.  Clearly, trigonometric parallaxes are
required to determine unambiguous distances to these nearby
dwarfs.

\section{Sample Characteristics and Preliminary Luminosity
Function} \label{sec:lf}

In this Paper we present new data on almost 300 objects.  These
data, combined with the previously known objects, represent 66\%
of the ultracool candidates in the 2MU2 sample.  The
color-magnitude and color-color diagrams for those data are shown
in Figure \ref{fig:done}.  In this section, we point out some of
the characteristics of the completed portion of the 2MU2 sample
that are relevant to the study of nearby low-mass stars and brown
dwarfs.

\begin{figure}
\epsscale{0.55} \plotone{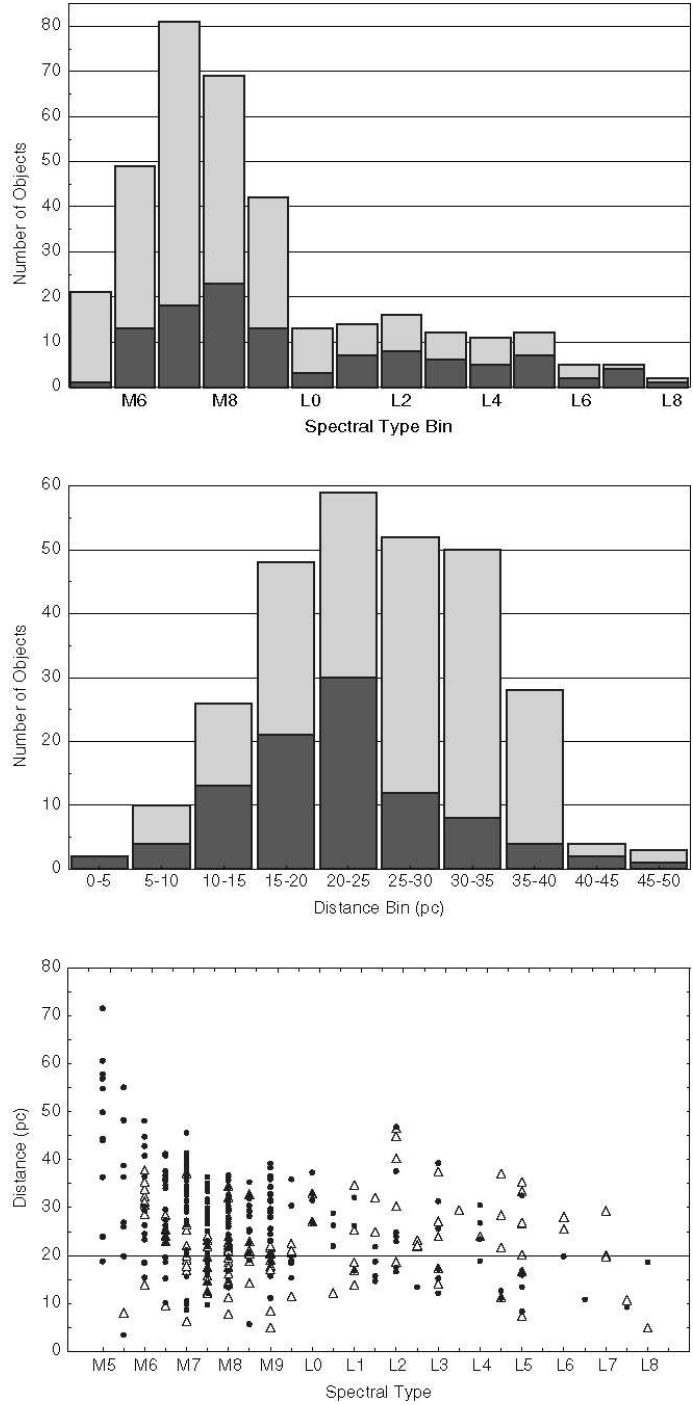}

\caption{Stacked histogram of the spectral type of all the dwarfs M5 and
later in our sample (\emph{top panel}) and distance distributions of the
M7--L8 dwarfs (\emph{middle panel}).  Darkly shaded region indicates
objects previously known while the lightly shaded region represents
additions.  \emph{Bottom panel} shows spectral type versus distance for
dwarfs later than M5.  Solid circles are new objects while open triangles
are previously known objects, and the solid horizontal line marks our
distance limit.  Multiplicity has been ignored in all three plots.}
\label{fig:hist}
\end{figure}

The upper panel of Figure \ref{fig:hist} shows the spectral type
distribution of the portion of the 2MU2 sample presented here,
separately identifying our new discoveries and previously known
ultracool dwarfs.  We have at least doubled the number of objects
in most spectral type bins, and tripled it in others.  Not until
L5 does the number of previously known objects outnumber our
additions.  Currently, the follow-up work is the most incomplete
for the faintest objects in the sample which is where we expect
to find the coolest, faintest objects, which have the latest
spectral types.

The distance distribution of the M7--L8 objects presented here is
shown in the middle panel of Figure \ref{fig:hist}.  While we
have not discovered any objects within five parsecs, we have
doubled the number of late-type dwarfs in every other distance
bin out to 40 pc.  In addition, the number of objects found
scales consistently with the increase in volume out to the 20 pc
bin.  The volume increases 3.4 times from 10 to 15 pc and the
number of objects we have found increases 3.2 times.  Similarly,
from 15-20 pc, the volume increases by a factor of 2.4 and the
number of objects included increases by 2.3.  This suggests that
we are indeed nearly complete out to 20 pc.

In the bottom panel of Figure \ref{fig:hist}, we show spectral
type versus distance.  Our current sample includes objects well
beyond the 20 pc limit even at spectral types later than L6.
This further supports that, once we have complete observations of
the full sample including the faintest candidates, we will have
identified a complete sample of late-M and L0--L8 dwarfs within
20 pc of the Sun.

\begin{figure}
\epsscale{0.55} \plotone{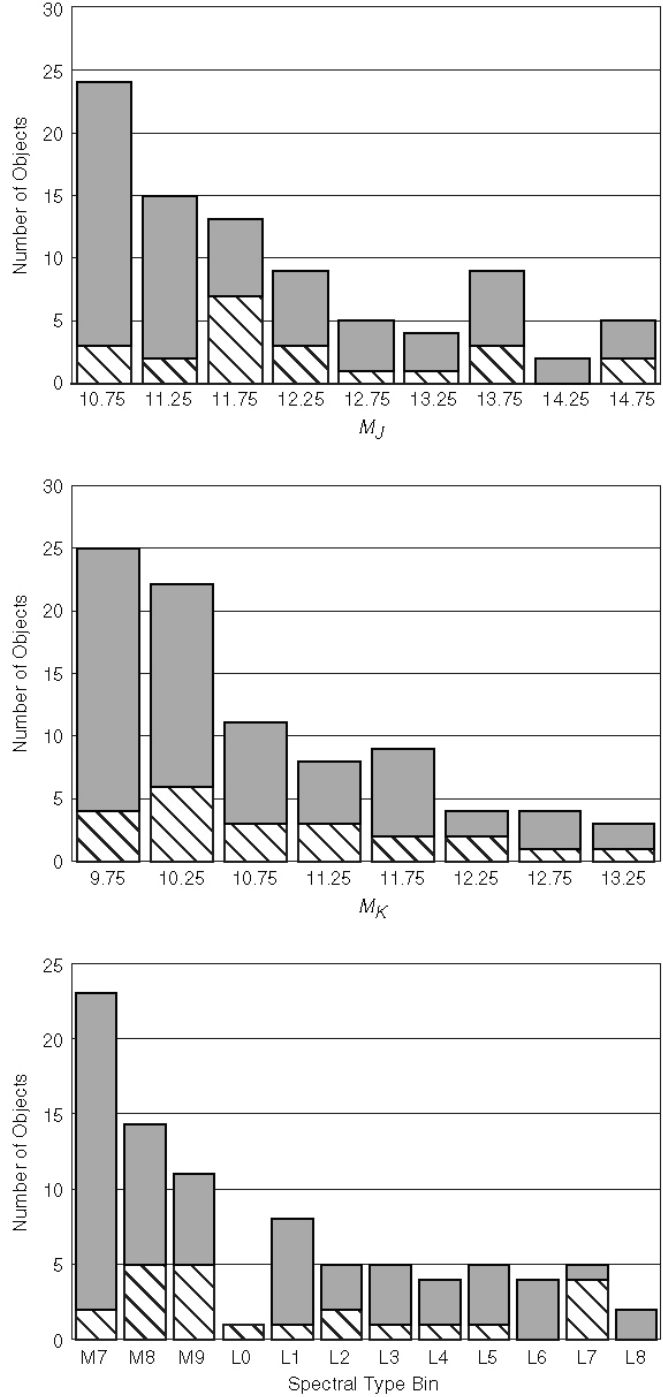}

\caption{Preliminary field luminosity function (\emph{top panel}), and
spectral type distribution (\emph{bottom panel}) for dwarfs within 20 pc.
Hatched region indicates objects with trigonometric parallaxes while
shading indicates $M_J$ is based on spectral type.  $M_J$ bins are 0.5
mags wide and are labeled with the centroid. Spectral type bins contain
both the integer type and the 0.5 class cooler subtype.  Multiplicity has
been taken into account and the magnitudes corrected.} \label{fig:lf}
\end{figure}

Preliminary field luminosity functions in $M_J$ and $M_{K_S}$
(based on the first results from our 2MU2 sample which has a
distance limit of 20 pc and an effective sky coverage of 40\%) are
shown in the top two panels of Figure \ref{fig:lf}.  The
components of known binaries are counted separately.  $M_J$ is
estimated using our spectral type/$M_J$ relation described in \S\
\ref{sec:cal} or using 2MASS $J$ band photometry and a
trigonometric distance estimate if available.  $M_{K_S}$ estimates
are obtained by subtracting the 2MASS $J-K_S$ color from $M_J$.

The faintest bins ($M_J>14$, $M_{K_S}>12.5$) should be regarded as
substantially incomplete.  The number of objects falls off at
magnitudes fainter than $M_J=11$ ($M_{K_S}=10.5$) as these bins
are dominated by evolving brown dwarfs rather than stable stars.
While stars remain at a given magnitude for billions of years,
brown dwarfs gradually cool to fainter and fainter magnitudes thus
depleting the brighter magnitude bins.

The peak at $M_J=13.75$ with 9 objects is not readily explained.
Since we are incomplete at these faint magnitudes ($M_J>13.5$)
and the number of objects in these bins will only go up with
increased completeness, it is possible that the peak is not an
isolated bump, but rather represents a change to an increasing
slope or a plateau of the luminosity function for $M_J>13.5$.
This could be explained as a population of old, massive brown
dwarfs that have evolved---essentially, this population has
migrated over time from $11.2<M_J<13.5$ to $M_J>13.5$.  On the
other hand, the peak is not evident in the $M_{K_S}$ luminosity
function.  It is important to keep in mind that these luminosity
functions are based on about two-thirds of our sample of
ultracool candidates, which is most lacking in the latest types.
While this feature is tantalizing and might have many
hypothetical explanations, a luminosity function based on our
complete sample, which includes objects down to $J=17$, will be
better suited to interpretation.

The spectral type distribution of the ultracool dwarfs presented
here with distance estimates within 20 pc is shown in the bottom
panel of Figure \ref{fig:lf}.  There is a striking lack of
L0s---the one object is actually a component of the 2M 0746
system.  One possibility for this is that the morphological
features corresponding to L0 span a smaller range of temperature
than the other types, thus objects evolve through L0 faster and we
see fewer objects.  However, we do not see a similar lack of L0s
in the entire sample (top panel of Figure \ref{fig:hist}).

\begin{figure}
\epsscale{1.0} \plotone{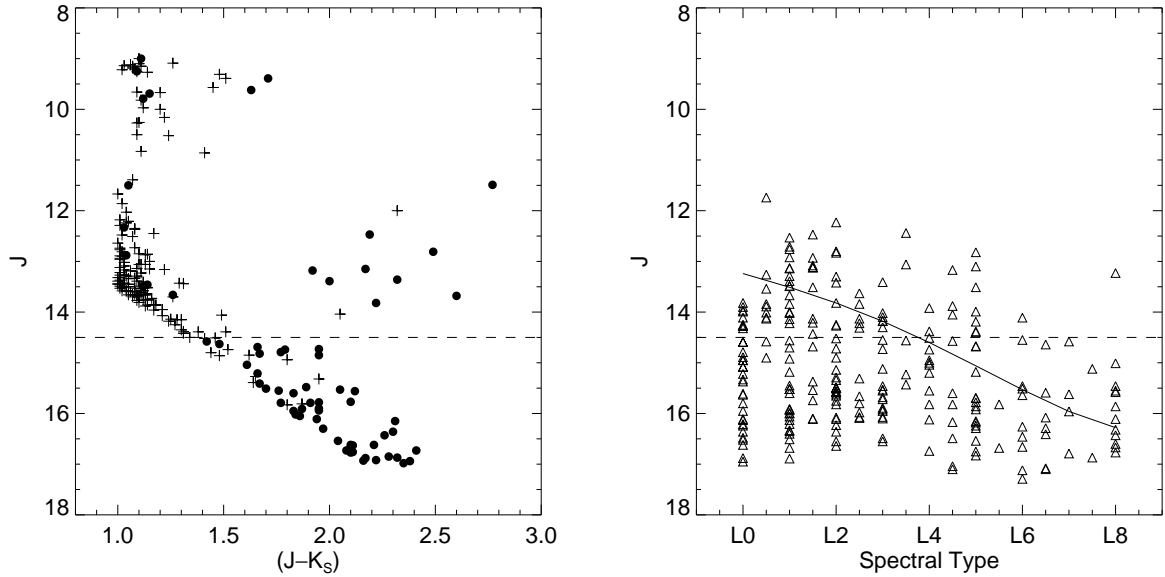}

\caption{Color-magnitude for the all of the targets in the sample that are
not presented here (\emph{left panel}) and $J$ versus spectral type for
all 250 objects listed in the L Dwarf Archive (\emph{right panel}).
Crosses are objects for which we have data while filled circles represent
objects that still require follow-up observations.  The dashed line shows
where our current incompleteness becomes significant and the solid line
marks the 20 pc limit.}\label{fig:left}
\end{figure}

The objects presented in this paper represent the brighter part of
our 2MU2 sample.  While we only present 66\% of the ultracool
candidates here, of the 219 remaining, we have obtained data on a
further 150 objects which will be discussed in a future paper.
Forty-eight of the 69 objects without follow-up observations are
faint and have proved elusive from 4 m class telescopes.  The left
panel of Figure \ref{fig:left} shows the color-magnitude
distribution of the objects not presented here, while the right
panel shows magnitude against spectral type for all of the objects
listed in the L Dwarf
Archive\footnote{\url{http://spider.ipac.caltech.edu/staff/davy/ARCHIVE/}}
along with a rough magnitude limit for our current sample and the
20 pc limit.  Based on these, we believe that many nearby
late-type L dwarfs are in our 2MU2 sample, but due to their
faintness, they still lack follow-up observations.  As discussed
above, these objects are essential to understanding the luminosity
and mass functions of ultracool dwarfs.

Based on initial data reduction of the 150 objects with follow-up
observations but not included in this paper, our preliminary luminosity
function is $\sim$90\% complete for the early-L dwarfs (L0--L4) found
within 20 pc. Taking into account our sky coverage of 40\%, we find a
early-L dwarf space density of 0.0019 pc$^{-3}$.

\section{Conclusions}
\label{sec:conclusions} We have described our project to create a
statistically robust, volume-limited survey for ultracool dwarfs
(spectral types M7--L8).  Our goal is to determine the infrared
luminosity function and constrain the mass function of late-type
stars and brown dwarfs.  We have presented our initial
findings---the discovery of 186 new late-M and L dwarfs.
Forty-seven of these are additions to the 20 pc nearby star census
of ultracool dwarfs, including two confirmed brown dwarfs.  These
data, combined with previously known nearby objects, are a
significant step towards estimating the mass distribution of
ultracool field dwarfs.  Our future work, especially with the
addition of the coolest objects, will illuminate even further the
statistical properties of low-mass stars and brown dwarfs in the
Solar neighborhood.

\acknowledgments

We would like to thank the numerous NOAO telescope operators and support
staff at Kitt Peak and Cerro Tololo that made this work possible and
endured our busy observing program: Alberto Alvarez, Ed Eastburn, Bill
Gillespie, Angel Guerra, Hal Halbedel, Hillary Mathis, and Sergio Pizarro.
We would also like to thank Nadya Gorlova and Curtis Cooper for assistance
at the 2001 July Kitt Peak observing run.  This research was partially
supported by a grant from the NASA/NSF NStars initiative, administered by
JPL, Pasadena, CA.  KLC acknowledges support from a NSF Graduate Research
Fellowship. This publication makes use of data products from the Two
Micron All Sky Survey, which is a joint project of the University of
Massachusetts and IPAC/CalTech, funded by NASA and the NSF; the NASA/IPAC
Infrared Science Archive, which is operated by JPL/CalTech, under contract
with NASA; and the SIMBAD database, operated at CDS, Strasbourg, France.
The Guide Star Catalog–II is a joint project of STScI and the Osservatorio
Astronomico di Torino.  STScI is operated by the AURA, for NASA under
contract NAS5-26555.  The participation of the Osservatorio Astronomico di
Torino is supported by the Italian Council for Research in Astronomy.
Additional support is provided by ESO, Space Telescope European
Coordinating Facility, the International GEMINI project and the ESA
Astrophysics Division.

\appendix

\section{The Brightest Sources}

\subsection{Introduction}

\begin{figure}
\epsscale{1.0}\plotone{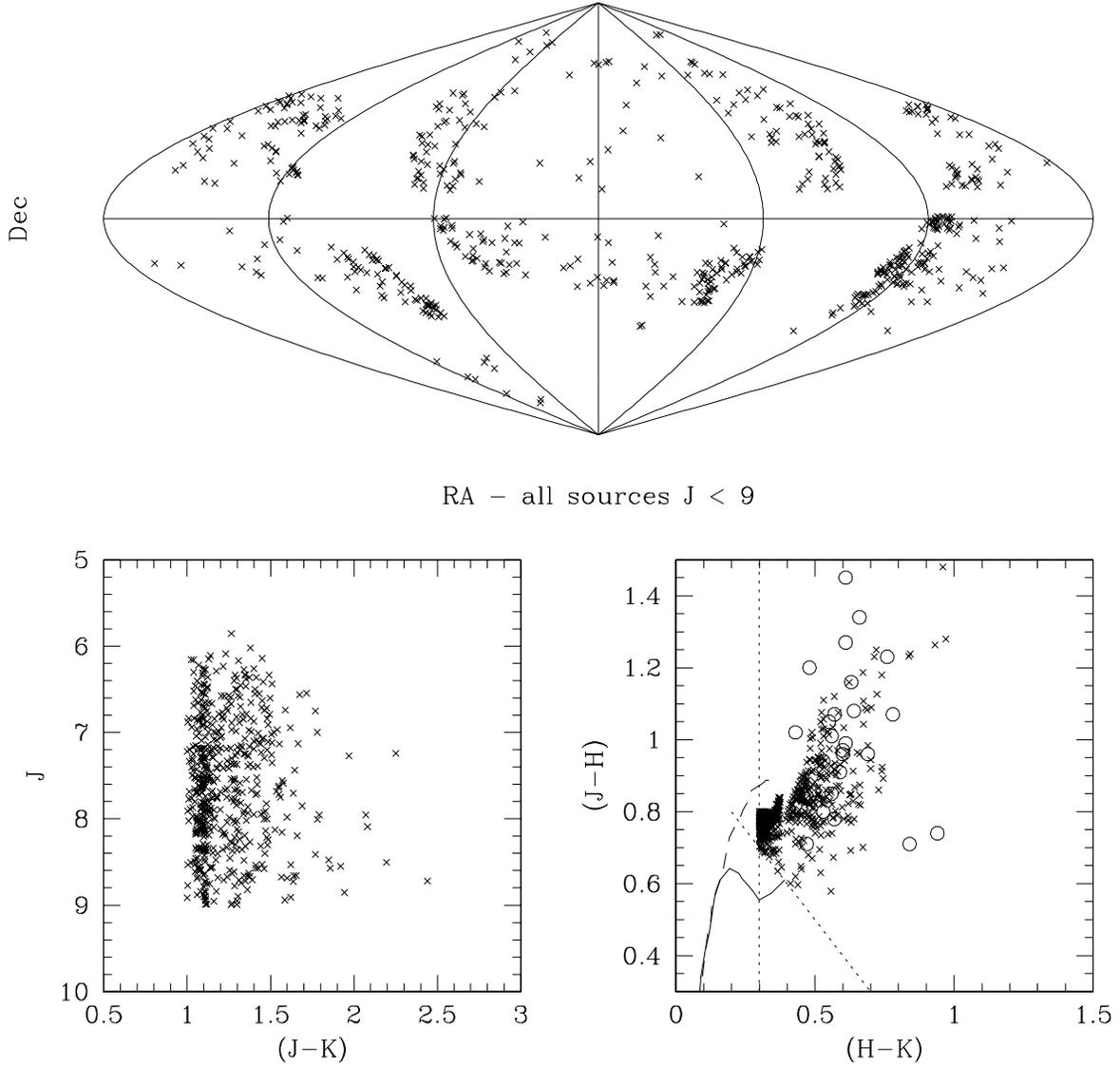}

\caption{The ($\alpha, \delta$) and near-infrared $J$/($J-K_S$) and
($J-H$)/($H-K_S$) distributions for all 588 sources in the 2MASS bright
ultracool sample.  Zero hours of RA is on the left and 12 hours is in the
center.  The dotted lines in the $JHK_S$ plane mark the selection
criteria; the dashed line plots the fiducial giant sequence; the solid
line marks the dwarf sequence; and the open circles plot data for L dwarfs
from \citet{K99}.}\label{fig:a1}
\end{figure}

As discussed in the text, 588 sources in our ultracool sample
have magnitudes brighter than $J=9$. Figure \ref{fig:a1} shows
the distribution of those stars on the celestial sphere and in
the near-infrared color-magnitude and color-color planes. Sources
with galactic latitudes in the range $-10\degr < b < 10\degr$
were excluded {\sl ab initio.} The fiducial dwarf and giant
sequences are plotted in the $JHK_S$ two-color diagram, together
with data for L dwarfs from \citet{K99}.  Note that we require
sources to have $(J-K_S)> 1.0$ and $(H-K_S)>0.3$; approximately
15\% of known L dwarfs have colors outside the limits bounded by
the present search criteria.

\subsection{Identifications}

\begin{figure}

\epsscale{1.0}\plotone{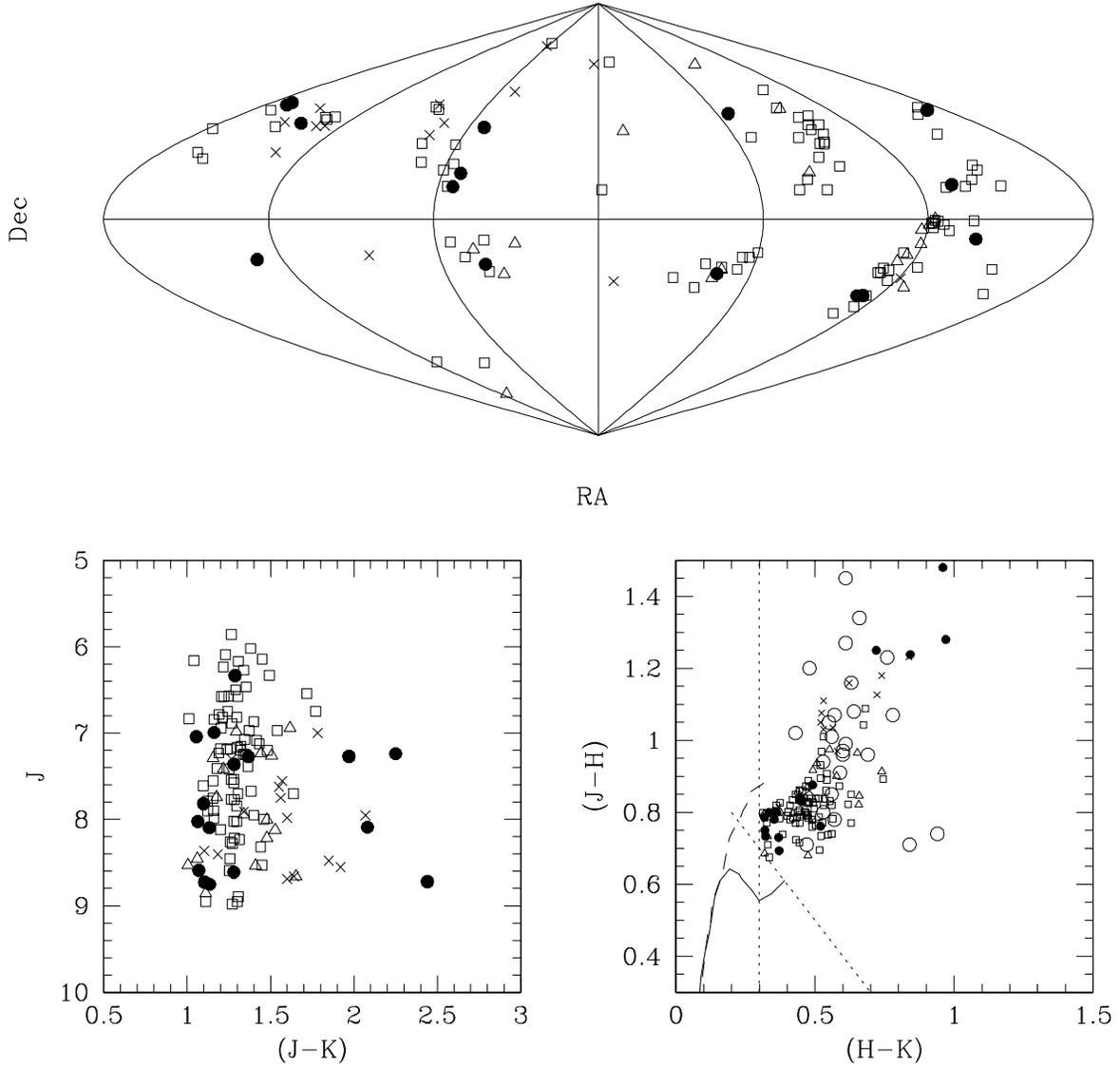}

\caption{The ($\alpha, \delta$) and near-infrared $J$/($J-K_S$) and
($J-H$)/($H-K_S$) distributions for 2MASS ultracool sources matched
against known carbon stars (crosses, Table \ref{tab:brightcarbon}), Miras
(open squares, Table \ref{tab:miras}), semi-regular variables (solid
points, Table \ref{tab:var}) and other late-type stars (open triangles,
Table \ref{tab:other}).  The lines of RA and the lines in the $JHK_S$
plane are the same as the previous figure.}\label{fig:a2}
\end{figure}

\begin{figure}
\epsscale{1.0} \plotone{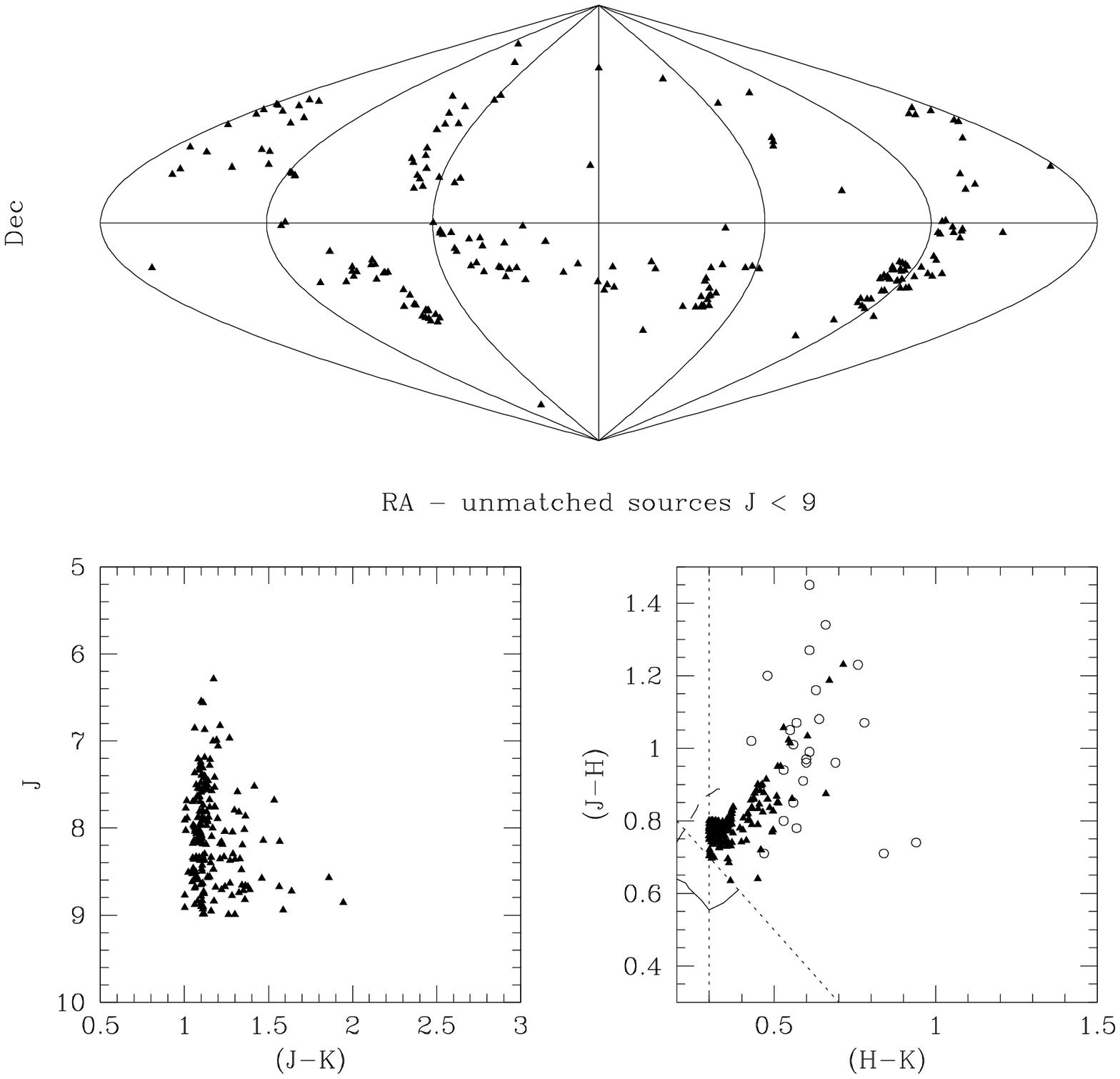}
\caption{The ($\alpha, \delta$) and near-infrared
$J$/($J-K_S$) and ($J-H$)/($H-K_S$) distributions for ultracool sources
with no counterpart listed by SIMBAD.  The lines of RA and the lines in
the $JHK_S$ plane are the same as the previous figure.}\label{fig:a3}
\end{figure}

All 588 sources were cross-referenced against the SIMBAD database
using a search radius of 2.0 arcminutes centered on the 2MASS
position. As discussed in the text, four stars are confirmed as
nearby dwarfs: G 180-11, G 139-3, Gl 866ABC and BD-01 3925D.  The
results for the remaining 584 candidates are as follows:

\begin{description}

\item[IRAS sources:] One hundred seventeen 2MASS candidates lie
within 10 to 15 arcseconds of a source from the IRAS catalog;
given the positional uncertainties of the IRAS astrometry, plus
the expectation that IRAS sources should be red in $(J-K_S)$,
these are highly likely to be the correct identification for the
2MASS source. The overwhelming majority of these sources are
expected to be dusty asymptotic giant branch stars (types M, S and
C) or red supergiants.  These sources are listed in Table
\ref{tab:iras}. We list $J$ magnitudes for extremely bright
sources where 2MASS $H$ and/or $K_S$ photometry is unavailable,
using ``:'' to denote uncertain measurements. Most of these
sources lie close to the Plane, as expected for young giants or
luminous AGB stars.

\item[Stellar sources:] One hundred thirty eight candidates lie
within 2 to 3 arcseconds of stars listed in either the Henry
Draper, Bonner Durchmusterung, Cape Durchmusterung, Cape
Photographic Durchmusterung, Guide Star Catalog, or PPM
catalogues. Data for these sources are given in Table
\ref{tab:stellar}. Most of the sample have colors close to the
giant sequence in the $JHK_S$ plane, with approximately fifteen
stars overlapping with the L dwarf distribution. The latter stars
are likely to be carbon stars.

\item[Carbon stars:] Thirteen stars in the sample are classed as
carbon stars in SIMBAD.  Figure \ref{fig:a2} shows their
distribution on the sky and in the near-infrared plane---there is
obvious overlap with the L dwarf distribution in the two-color
diagram. Data for these stars are listed in Table
\ref{tab:brightcarbon}.

\item[Mira variables:] Eighty stars are identified as
Miras---M-type long-period variables.  Most of these stars lie
close to the Galactic Plane, with a particular concentration in
the ScoCen region (towards the Bulge). Data are listed in Table
\ref{tab:miras}, and plotted in Figure \ref{fig:a2}.

\item[Semi-regular variables:] Sixteen stars are identified as
semi-regular (AGB/RGB) variables. As might be expected, the
spatial and color-magnitude distributions are similar to those of
the Miras (Figure \ref{fig:a2}, Table \ref{tab:var}), with a
strong concentration towards the Plane and the Bulge.

\item[Other:] A further eighteen stars are identified as
late-type stars based on cross-referencing against SIMBAD.  Those
stars are listed in Table \ref{tab:other}.  Most are late-type
giants. In particular,  StM 218 is from Stephenson's survey of
high-latitude red giants \citep{stephenson}, with additional
spectroscopy by \citet{sharples}; BR B0954-0947 is from the APM
QSO survey \citep{KHI}; and the DENIS source, from \citet{NN4} is
a known red-giant variable.  Of the remaining stars, WOH S 11 is
listed as type M by SIMBAD, but with no additional information,
and TX CVn is a symbiotic binary.

\item[Unmatched sources:] Two hundred two sources have no obvious
counterpart in the SIMBAD database. 2MASS data for those sources
are listed in Table \ref{tab:unmatched}, and the spatial and
color-magnitude distribution plotted in Figure \ref{fig:a3}.  As
discussed in the text, all of these sources have optical
counterparts, indicating low proper motions, and most have
optical/near-infrared colors which are inconsistent with late-type
dwarfs. Given those characteristics, plus the distribution in
Galactic coordinates, all are likely to be pre-main-sequence stars
or AGB stars.

\end{description}

\subsection{Discussion}

\subsubsection{The Reddest Candidates}

The overwhelming majority of the sources listed in Tables
\ref{tab:iras} to \ref{tab:unmatched} are clearly red giants.
Nonetheless, eight sources in Table \ref{tab:stellar} and
twenty-three sources in Table \ref{tab:unmatched} have
($H-K_S)>0.45$---colors potentially consistent with L dwarfs. We
discuss these sources in detail below:

\begin{itemize}

\item The stellar candidates: all have excellent positional
agreement between 2MASS and SIMBAD (ICRS) co-ordinates. Three
(V355 Gem, CD CVn and BD-08 2741) are known giant stars, and it
is extremely likely that the remaining eight are also red giants.

\begin{description}

\item[2MASSI J0609248+773327:] BD+77 225, spectral type K0:
$V=9.49$, so ($V-K_S)=3.47$, consistent with $M_V \sim 9.0$ and
spectral type M0 if it were a dwarf. Those parameters would place
it within 10 parsecs of the Sun. However, the low proper motion
(($\mu_{\alpha},\mu_{\delta})=-12, -0.2$ mas yr$^{-1}$) and the
inconsistency between the ($V-K_S$) color and the observed
spectral type indicates that this star is a giant.

\item[2MASSI J0700365+260818:] GSC 01899-00620, and also
identified as V355 Gem and IRAS 06575+2612.  This is clearly an
evolved star.

\item[2MASSI J1300025+472632:] CD CVn, listed as $V=9.39$,
($B-V)=1.19$, spectral type K0 III and  $\pi = 3.31\pm1.21$ by
SIMBAD. This is a red giant variable, with ($V-K_S)=3.28$.

\item[2MASSI J1341018+563452:] HD 238271, $V=9.55$, ($B-V)=1.51$,
spectral type K5, negligible proper  motions (-7, 6.2 mas
yr$^{-1}$). The low proper motions confirm the star as a red
giant, ($V-K_S)=3.64$.

\item[2MASSI J1724215+652915:] BD+65 1182, $V=9.61$, ($B-V)=1.45$,
($V-K_S)=3.45$, spectral type K2, negligible proper motions (-3,
-6 mas yr$^{-1}$). Again, the low motions, measured colors and
observed spectral type are most consistent with a giant.

\item[2MASSI J1726192+601748:] BD+60 1757, $V=9.94$, ($B-V)=1.66$,
($V-K_S)=4.15$, spectral type K2, negligible motions (-12, 9 mas
yr$^{-1}$). As with 2M 1724, the low motions, measured colors and
observed spectral type are most consistent with a giant.

\item[2MASSI J1816212+202817:] HD 348183, $V=9.06$, ($B-V)=1.52$,
negligible motions (-7, -6 mas yr$^{-1}$), spectral type K7.
($V-K_S)=3.92$, and another red giant.

\item[2MASSI J2015149$-$153626:] GSC 06315-00584, also identified
as NSV 12940, a red giant variable.

\end{description}

\item The 34 sources with no SIMBAD identification: as noted
above, all of these sources have optical counterparts. Genuine
ultracool dwarfs are expected to have ($B-R)>4$ and ($R-K_S)>7$
for ($H-K_S)>0.45$ (spectral type later than M8). The
corresponding distances (for ultracool dwarfs) are less than 5
parsecs, so the absence of any measured proper motion ($\mu < 0.1$
arcsec yr$^{-1}$) would require transverse velocities of less than
2.5 kms$^{-2}$.

\begin{description}

\item[2MASSI J0547281$-$214723:] ($B-R)_{USNO}=3.0$,
($R-K_S)=4.2$. Spectroscopy with the CTIO 1.5 m confirms this as a
carbon star.

\item[2MASSI J0613450+522540:]  ($B-R)_{USNO}=2.1$, ($R-K_S)=6.5$,
with no evidence for motion between POSS I and either POSS II or
2MASS. The source appears significantly brighter on the POSS II
$F$ plate than on the POSS I $E$, suggesting variability, and
identify this star as a probable AGB variable.

\item[2MASSI J0620521$-$164541:]  ($B-R)_{USNO}=1.2$,
($R-K_S)=7.2$, with no evidence for significant proper motion.
CTIO spectroscopy confirms this as an M giant.

\item[2MASSI J0641403$-$282102:]  ($B-R$)$_{USNO}=0.8$,
($R-K_S)=5.8$, with no evidence for significant proper motion.
CTIO spectroscopy confirms this as an M giant.

\item[2MASSI J0650548$-$372922:]  ($B-R$)$_{USNO}=4.1$,
($R-K_S)=7.9$, with no evidence for significant proper motion.
CTIO spectroscopy confirms this as an M giant.

\item[2MASSI J0652228+452045:]  ($B-R$)$_{USNO}=1.6$,
($R-K_S)=5.2$, and no evidence for motion between POSS I and
either POSS II or 2MASS. This is likely to be a red giant.

\item[2MASSI J0657520+662111:]  ($B-R$)$_{USNO}=1.3$,
($R-K_S)=5.3$, and no evidence for motion between POSS I and
either POSS II or 2MASS. Given the Galactic latitude,
$b=+25\degr$, the colors are most consistent with a reddened
background star.

\item[2MASSI J0658118+263535:]  ($B-R$)$_{USNO}=5.0$,
($R-K_S)=4.7$, with no evidence for significant proper motion.
CTIO spectroscopy confirms this as an M giant.

\item[2MASSI J0701322$-$381421:]  ($B-R$)$_{USNO}=3.2$,
($R-K_S)=5.4$, with no evidence for significant proper motion.
CTIO spectroscopy confirms this as an M giant.

\item[2MASSI J0703356$-$404748:]  ($B-R$)$_{USNO}=5.9$,
($R-K_S)=5.5$, $b=-15\degr$. Highly likely to be a reddened
source.

\item[2MASSI J0710483+305546:] ($B-R$)$_{USNO}=4.7$,
($R-K_S)=4.4$. No motion evident, and clearly fainter on POSS II
than POSS I. Likely to be a red giant variable.

\item[2MASSI J0710574+475818:] ($B-R$)$_{USNO}=3.8$,
($R-K_S)=5.0$, and no evidence for significant proper motion.
Likely to be a red giant or reddened background star.

\item[2MASSI J0721404+194350:]  ($B-R$)$_{USNO}=1.9$,
($R-K_S)=4.9$. CTIO spectroscopy identifies this as an M giant.

\item[2MASSI J0813343$-$051321:]  ($B-R$)$_{USNO}=4.0$,
($R-K_S)=4.9$. CTIO spectroscopy identifies this as a carbon star.

\item[2MASSI J0829151+182307:]  ($B-R$)$_{USNO}=4.6$,
($R-K_S)=4.4$. No evidence for motion; colors strongly suggest
highly reddened object.

\item[2MASSI J1010015$-$023743:]   ($B-R$)$_{USNO}=3.1$,
($R-K_S)=4.4$. CTIO spectroscopy identifies this high-latitude
($b=41\degr$) candidate as an M giant.

\item[2MASSI J1158169$-$253753:]  ($B-R$)$_{USNO}=1.6$,
($R-K_S)=4.3$. CTIO spectroscopy identifies this as another
high-latitude ($b=36\degr$) M giant.

\item[2MASSI J1502099+593121:]  ($B-R$)$_{USNO}=2.2$,
($R-K_S)=6.4$, but relatively faint on the POSS II IVN plate
($I\sim15$) and barely visible on the IIIaJ plate ($B\sim21$).
This is likely to be a high latitude red giant variable.

\item[2MASSI J1502582$-$355111:]   ($B-R$)$_{USNO}=0.8$,
($R-K_S)=6.8$. Significantly brighter on the UKST IIIaF plate than
on the POSS I 103aE plate, and no evidence for motion. This is
likely to be a red giant variable.

\item[2MASSI J1930155$-$232048:] ($B-R$)$_{USNO}=0.9$,
($R-K_S)=8.2$, no evidence for motion and near the Lupus dark
cloud.  Likely to be a pre-main sequence star or dusty giant.

\item[2MASSI J1942441$-$295436:]  ($B-R$)$_{USNO}=4.6$,
($R-K_S)=4.9$, and no evidence for motion. Likely to be a dusty
giant.

\item[2MASSI J2025464$-$163148:] ($B-R$)$_{USNO}=0.2$,
($R-K_S)=6.3$, and no evidence for motion POSS I/POSS II/2MASS.
Brighter on second epoch UKST IIIaF than IVN, suggesting
identification as a red giant variable.

\item[2MASSI J2044540$-$074359:] ($B-R$)$_{USNO}=2.1$,
($R-K_S)=6.8$, and no evidence for motion.  Significantly fainter
on UKST IIIaF plate, strongly suggesting variability and
identification as a red giant.

\end{description}
\end{itemize}

\subsubsection{Cross-checks Against Existing Catalogs}

As an additional test, all of the bright ultracool candidates were
cross-referenced against the third Catalog of Nearby Stars
\citep[pCNS3]{pCNS3} and against Luyten's NLTT proper motion
catalog \cite{nltt}.  The two catalogs were cross-referenced using
a search based on position with a search radius of 180\arcsec.
There are only twelve matches, including G180-11, G139-3, BD-01
3925D, and EZ Aqr (Gl 866ABC) which are discussed in the main
text.  In each of the remaining nine cases, it is clear that the
2MASS ultracool source is not the NLTT star.

\subsection {Conclusions}

The overwhelming majority of the 588 sources in the ultracool
sample with $J<9$ can be eliminated as candidate nearby dwarfs:
386 have previously-cataloged optical or infrared counterparts,
and the overwhelming majority of those are AGB stars.  Of the
remaining 202 sources, only twenty three have colors sufficiently
red to be candidate L dwarfs. All of the latter are visible on
photographic plate material, and none have either measurable
proper motions or optical/near-infrared colors consistent with
ultracool dwarfs. We conclude that four proper-motion objects, G
180-11, G139-3, BD-01 3925D, and Gl 866 (Table \ref{tab:bright})
are the only genuine late-type dwarfs amongst the ultracool
candidates with $J\le9$.

\clearpage

\clearpage

\clearpage

\clearpage

\clearpage

\clearpage

\clearpage

\clearpage

\clearpage

\clearpage

\clearpage

\clearpage

\clearpage

\clearpage

\clearpage

\clearpage

\begin{deluxetable}{lr}
\tablewidth{0pt}\tablecaption{Accounting of Steps to Create the
2MU2 Sample\label{tab:cuts}} \tablehead{\colhead{Item} &
\colhead{Number} } \startdata
\bf{2MASS Hits} & \bf{11,316,668}\\
\hline
Clouds & 1,271,555\\
LMC, SMC, \& 47 Tuc & 383,472\\
M31 \& M33 & 10,292\\
$J$, ($J-K$) & 9,414,914\\
$(J-H)$, ($H-K$) & 227,904\\
$R$, ($R-J$) & 5337\\
Giants&698\\
Flags& 310\\
Reddening Regions & 514\\
\hline \hline
\bf{Total}& \bf{1672}\\
\enddata
\tablecomments{The sample started as $\approx$11 million 2MASS
hits. The following (non-bold face) entries list reasons for
eliminating objects from the sample and their contribution to
reducing the final sample size to 1672 candidates.  An explanation
of each reason for elimination is given in the text.}
\end{deluxetable}

\begin{deluxetable}{l@{}c@{}llrr}
\tablewidth{0pt}\tablecaption{Cataloged Clouds\label{tab:red1}}
\tablehead{\colhead{Region} & \colhead{} & \colhead{$l_{min}$} &
\colhead{$l_{max}$} & \colhead{$b_{min}$} & \colhead{$b_{max}$}}
\startdata
Cepheus     &  & 100       & 120       & \phs11 &  \phs22 \\
\hline
Per OB2     &  & 154       & 163       & $-$25  & $-$10 \\
Taurus      &  & 163       & 202.5     & $-$22  & $-$10 \\
Orion B     &  & 202.5     & 210       & $-$21  & $-$10 \\
Orion A     &  & 210       & 218       & $-$21  & $-$14 \\
Mon R2      &  & 210       & 218       & $-$14  & $-$10 \\
\hline
Cham        &  & 295       & 305       & $-$20  & $-$12 \\
\hline
Lupus       &  & 333       & 350       & \phs10 & \phs22 \\
$\rho$ Oph  &  $ \left\{ \begin{array}{@{}l@{}} \\ \\ \end{array}
\right. $ & $ \begin{array}{@{}l@{}} 350\\ 350\\ \end{array} $ & $
\begin{array}{@{}l@{}} \phn\phn2 \\ \phn13 \\ \end{array} $ & $
\begin{array}{@{}r@{}} \phn13 \\ \phn10  \end{array} $ & $
\begin{array}{@{}r@{}} \phn24 \\ \phn19  \end{array} $ \\
Upper Sco   &  & 355       & 363       & \phs16 & \phs27 \\
\hline R CrA & $ \left\{ \begin{array}{@{}l@{}} \\ \\ \end{array}
\right. $ & $ \begin{array}{@{}l@{}} \phn\phn4\\ 352\\
\end{array} $ & $ \begin{array}{@{}l@{}} \phn15\\ \phn\phn4\\
\end{array} $ & $ \begin{array}{@{}r@{}} -15\\ -22\\ \end{array}
$ & $ \begin{array}{@{}r@{}} -10\\ -20\\ \end{array} $ \\
\enddata
\tablecomments{These regions have been eliminated from the sample.
The positions have been expanded upon from regions listed by
\citet{Dame} and \citet{CDN}.  Some of the regions overlap.}
\end{deluxetable}

\begin{deluxetable}{rrrr}
\tablewidth{0pt}\tablecaption{Uncataloged Reddening
Regions\label{tab:red2}} \tablehead{\colhead{$l_{min}$} &
\colhead{$l_{max}$} & \colhead{$b_{min}$} & \colhead{$b_{max}$} }
\startdata
0   &96  & $-$16  & \phs16 \\
150 &180 & \phs10 & \phs13 \\
180 &360 & $-$13  & \phs13 \\
199 &214 & $-$13  & $-$27  \\
308 &310 & \phs13 & \phs16 \\
\enddata
\tablecomments{Objects lying in these regions have been removed
from the sample.}
\end{deluxetable}

\begin{deluxetable}{lr}
\tablewidth{0pt}\tablecaption{Accounting of Steps to Refine $J>9$
Portion of the 2MU2 Sample\label{tab:morecuts}}
\tablehead{\colhead{Item} & \colhead{Number} } \startdata
\bf{Faint Candidates} & \bf{1084}\\
\hline ($F-J$)/($J-K$) & 137\\
Visual Inspection & 211\\
SIMBAD & 24\\
Clouds & 41\\
Bland & 34\\
\hline \hline
\bf{Ultracool Candidates}& \bf{637}\\
\enddata
\tablecomments{Similar to Table \ref{tab:cuts} except listing the
cuts applied only to the fainter ($J>9$) portion of the sample and
their contribution to reduce the number of ultracool candidates to
637.}
\end{deluxetable}

\pagestyle{empty}
\begin{deluxetable}{llrrrlccccccc}
\tablewidth{0pt}
\tabletypesize{\scriptsize}
\rotate
\tablecaption{Previously Known Cool Dwarfs Recovered in the 2MU2 Sample\label{tab:known}}
\tablehead{ \colhead{2MASSI} & \colhead{Other} & \colhead{$J$} & \colhead{$J-H$} &
\colhead{$J-K_S$} & \colhead{Spectral}& \colhead{$M_J$}& \colhead{$d$}& \colhead{$M_{J_{\pi}}$}& \colhead{$d_{\pi}$}& \colhead{$\pi$}& \colhead{Mult.} & \colhead{Discovery}\\
\colhead{Designation} & \colhead{Names} & \colhead{} & \colhead{} &
\colhead{} & \colhead{Type}& \colhead{}& \colhead{(pc)}& \colhead{}& \colhead{(pc)}& \colhead{Ref.}& \colhead{Ref.} & \colhead{Ref.}
}
\startdata
0024246$-$015819$^{*}$  & BRI 0021$-$0214                 & 11.860\tablenotemark{a} & 0.740 & 1.280& M9.5                  & 11.6$\pm$0.1                  & 11.3$\pm$0.7                  & 11.56$\pm$0.10                  & 11.5$\pm$0.5                    & 1  &           & 1   \\
0024442$-$270825        & LHS 1070/LP 881- 64             & 9.262                   & 0.730 & 1.030& M5.5                  &  \nodata                      &  \nodata                      & 9.92$\pm$0.20                   & 7.39$\pm$0.66                   & 2  &           & 2   \\
0027559$+$221932$^{*}$  & LP 349- 25                      & 10.608                  & 0.638 & 1.047& M8                    & 11.2$\pm$0.2                  & 7.80$\pm$0.6                  &                                 &                                 &    &           & 3   \\
0030300$-$145033        &                                 & 16.792                  & 1.437 & 2.414& L7                    & 14.4$\pm$0.2                  & 29.4$\pm$3.4                  &                                 &                                 &    &           & 4   \\
0051107$-$154417        &                                 & 15.230                  & 1.082 & 1.806& L3.5                  & 12.9$\pm$0.2                  & 29.5$\pm$3.0                  &                                 &                                 &    &           & 4   \\
0052546$-$270559        & RG 0050.5$-$2722                & 13.611                  & 0.617 & 1.072& M8                    & 11.2$\pm$0.2                  & 31.0$\pm$2.6                  & 11.94$\pm$0.47                  & 21.6$\pm$4.7                    & 3  &           & 5   \\
0058425$-$065123        &                                 & 14.320                  & 0.875 & 1.413& L0                    & 11.7$\pm$0.1                  & 33.0$\pm$2.1                  &                                 &                                 &    &           & 4   \\
\enddata
\end{deluxetable}

\pagestyle{headings}
\begin{deluxetable}{rrrrrrrrl}
\tabletypesize{\scriptsize} \tablecolumns{9} \tablewidth{0pt}
\tablecaption{Late-type Dwarfs with $J<9$\label{tab:bright}}
\tablehead{\colhead{$\alpha$ (2000)} & \colhead
{$\delta$} & \colhead {$V$} & \colhead{$J$} & \colhead{$J-H$} &
\colhead {$J-K_S$} &  \colhead{ID}} \startdata
15 55 31.8 &    35 12 02 & 13.7 & 8.999 & 0.709 & 1.013 & G 180-11, M4.5      \\
16 58 25.3 &    13 58 10 & 13.1 & 8.859 & 0.575 & 1.122 & G 139-3, M4         \\
20 13 55.9 & $-$00 52 53 & 10.5 & 8.206 & 0.764 & 1.101 & BD-01 3925D         \\
22 38 33.3 & $-$15 18 02 & 12.2 & 6.577 & 0.612 & 1.010 & EZ Aqr, Gl 866, M5.5\\
\enddata
\end{deluxetable}

\pagestyle{empty}
\begin{deluxetable}{llllllllccc}
\tablewidth{0pt} \tabletypesize{\scriptsize} \rotate
\tablecaption{M7--L8 Dwarfs Discovered Within 20
pc\label{tab:latenear}} \tablehead{ \colhead{2MASSI} &
\colhead{Other} & \colhead{$J$} & \colhead{$J-H$} &
\colhead{$J-K_S$} & \colhead{Obs.} & \colhead{Telescope} &
\colhead{Spectral}& \colhead{$M_J$}& \colhead{$d$}& \colhead{Other}\\
\colhead{Designation} & \colhead{Names} & \colhead{} & \colhead{}
& \colhead{} & \colhead{Date} & \colhead{} & \colhead{Type}&
\colhead{}& \colhead{(pc)}& \colhead{Ref.} } \startdata
0019262$+$461407 &                            & 12.609 & 0.676  & 1.135 &2000 Oct 02 & KP 2.1m & M8                      & 11.2$\pm$0.2                    & 19.5$\pm$1.6                   &       \\
0019457$+$521317 &                            & 12.820 & 0.748  & 1.204 &2000 Oct 02 & KP 2.1m & M9                      & 11.5$\pm$0.1                    & 18.7$\pm$1.2                   &       \\
0109511$-$034326 & LP 647- 13                 & 11.695 & 0.774  & 1.277 &2000 Sep 29 & KP 2.1m & M9                      & 11.5$\pm$0.1                    & 11.1$\pm$0.7                   & 1     \\
0141032$+$180450 &                            & 13.822 & 0.772  & 1.315 & \nodata    & \nodata & L4.5\tablenotemark{a}   & 13.3$\pm$0.5                    & 12.6$\pm$2.7                   & 2     \\
0144353$-$071614 &                            & 14.187 & 1.183  & 1.904 &2002 Jan 24 & CT 4m   & L5                      & 13.6$\pm$0.2                    & 13.4$\pm$1.5                   & 3,4   \\
0148386$-$302439 &                            & 12.282 & 0.641  & 1.038 &2002 Jan 28 & CT 1.5m & M7.5                    & 11.0$\pm$0.2                    & 18.4$\pm$1.8                   &       \\
0213288$+$444445 &                            & 13.512 & 0.740  & 1.269 &2001 Jul 23 & KP 4m   & L1.5                    & 12.1$\pm$0.2                    & 18.7$\pm$1.4                   &       \\
0251148$-$035245 &                            & 13.082 & 0.821  & 1.429 &2002 Jan 23 & CT 4m   & L3                      & 12.7$\pm$0.2                    & 12.1$\pm$1.1                   & 2     \\
\enddata

\tablecomments{$M_J$ and distances listed are estimates obtained
using our calibration.  Data for multiple systems are based on the
system's combined spectral type and photometry and are marked with
parentheses.}

\tablenotetext{a}{ST from \citet{wilson}}
\tablenotetext{b}{ST change from M6 in Paper III}
\tablenotetext{c}{ST T0 in NIR \citet{Geballe}}
\tablenotetext{d}{Binary with $M_J=10.96$ and 12.06 (L. Close, in preparation). $d=11.1$ pc.}
\tablenotetext{e}{ST change from M6 in Paper III}
\tablenotetext{f}{Has \ion{Li}{1} absorption.  See \S\ \ref{sec:li}}
\tablenotetext{g}{Has strong H$\alpha$ emission.  See \S\ \ref{sec:active}}
\tablenotetext{h}{Low galactic latitude and is either a late-L dwarf or a reddened background star.}
\tablenotetext{i}{ST from \citet{sdss1}}
\tablenotetext{j}{ST from \citet{wilson}}
\tablenotetext{k}{$d_{\pi} = 5.67\pm0.02$ (Paper IV)}
\tablenotetext{l}{Has \ion{Li}{1} absorption and H$\alpha$ emission.  See \S\ \ref{sec:li}}
\tablenotetext{m}{$d_{\pi} = 18.9\pm0.7$ \citep{hipparcos}}

\tablerefs{(1) Paper III; (2) \citet{wilson}; (3) \citet{liebert03}; (4) \citet{Kendall}
(5) \citet{SDSS2}; (6) \citet{Geballe}; (7) \citet{Hawley}; (8) \citet{mccaughrean};
(9) \citet{NN5}; (10) \citet{twhya}; (11) \citet{ceso}; (12) \citet{sdss1}; (13) Paper IV;
(14) \citet{LSR1}; (15) \citet{K01}}

\end{deluxetable}

\pagestyle{empty}
\begin{deluxetable}{llllllllllc}
\tablewidth{0pt} \tabletypesize{\tiny} \rotate
\tablecaption{M7--L8 Dwarfs Discovered Outside 20
pc\label{tab:latefar}} \tablehead{ \colhead{2MASSI} &
\colhead{Other} & \colhead{$J$} & \colhead{$J-H$} &
\colhead{$J-K_S$} & \colhead{Obs.} & \colhead{Telescope} &
\colhead{Spectral}&
\colhead{$M_J$}& \colhead{$d$}& \colhead{Other}\\
\colhead{Designation} & \colhead{Names} & \colhead{} & \colhead{}
& \colhead{} & \colhead{Date} & \colhead{} & \colhead{Type}&
\colhead{}& \colhead{(pc)}& \colhead{Ref.} } \startdata
0021508$+$422050 &                 & 14.283 & 0.774  & 1.310 &2002 Jan 25 & KP 4m   & M9?                   & 11.5$\pm$0.3 & 36.6$\pm$4.7                 &     \\
0022137$+$120305 &                 & 13.469 & 0.661  & 1.023 &2001 Jul 21 & KP 4m   & M7.5                  & 11.0$\pm$0.2 & 31.7$\pm$3.1                 &     \\
0025576$+$391136 &                 & 13.565 & 0.654  & 1.084 &2002 Jan 25 & KP 4m   & M8                    & 11.2$\pm$0.2 & 30.3$\pm$2.5                 &     \\
0036448$-$203131 &                 & 14.383 & 0.710  & 1.309 &2002 Jan 25 & CT 4m   & M9?                   & 11.5$\pm$0.3 & 38.3$\pm$4.9                 &     \\
0049267$-$063546 &                 & 13.347 & 0.779  & 1.246 &2002 Jan 26 & CT 4m   & M8.5                  & 11.3$\pm$0.2 & 25.4$\pm$1.8                 &     \\
0055294$+$511536 &                 & 13.465 & 0.738  & 1.167 &2001 Jul 16 & KP 4m   & M7.5                  & 11.0$\pm$0.2 & 31.6$\pm$3.1                 &     \\
0107042$+$243527 &                 & 13.507 & 0.770  & 1.288 &2001 Jul 14 & KP 4m   & M9                    & 11.5$\pm$0.1 & 25.6$\pm$1.7                 & 1   \\
0117474$-$340325 &                 & 15.184 & 0.970  & 1.700 &2002 Jan 25 & CT 4m   & L2?                   & 12.3$\pm$0.3 & 37.6$\pm$5.9                 &     \\
\enddata

\tablecomments{$M_J$ and distances listed are estimates obtained
using our calibration.  Data for multiple systems are based on the
system's combined spectral type and photometry and are marked with
parentheses.}

\tablenotetext{a}{ST change from M6 in Paper III}
\tablenotetext{b}{ST from \citet{Kendall}}
\tablenotetext{c}{Data listed are based on photometric distance
from \citet{NN5}.}
\tablenotetext{d}{ST from \citet{twhya}}
\tablenotetext{e}{Data listed are based on photometric distance
from \citet{tinney95}.}
\tablenotetext{f}{Displays strong H$\alpha$ emission.  See \S\ \ref{sec:active}}
\tablenotetext{g}{ST from \citet{Hawley}}
\tablenotetext{h}{Displays strong H$\alpha$ emission.  See \S\
\ref{sec:active}}

\tablerefs{(1) \citet{wilson}; (2) \citet{NN5};  (3) \citet{Luyten1936}; (4) Paper III;
(5) \citet{Lodieu}; (6) \citet{Kendall}; (7) \citet{Bouy}; (8) \citet{NN4};
(9) \citet{twhya}; (10) \citet{tinney95}; (11) \citet{G03}; (12) \citet{Hawley}}
\end{deluxetable}

\pagestyle{empty}
\begin{deluxetable}{llllllllllc}
\tablewidth{0pt} \tabletypesize{\scriptsize} \rotate
\tablecaption{Early-type M Dwarfs Discovered Within 20
pc\label{tab:earlynear}} \tablehead{ \colhead{2MASSI} &
\colhead{Other} & \colhead{$J$} & \colhead{$J-H$} &
\colhead{$J-K_S$} & \colhead{Obs.} & \colhead{Telescope} & \colhead{Spectral}&
\colhead{$M_J$}& \colhead{$d$}& \colhead{Other}\\
\colhead{Designation} & \colhead{Names} & \colhead{} & \colhead{}
& \colhead{} & \colhead{Date} & \colhead{} & \colhead{Type}&
\colhead{}& \colhead{(pc)}& \colhead{Ref.} } \startdata
0035441$-$054110 & LP 645- 53          & 10.717 & 0.633  & 1.001 &2000 Sep 30 & KP 2.1m & M5                    & \phn9.4$\pm$0.3 & 18.7$\pm$2.3  & 1     \\
0214125$-$035743 & LHS 1363/LP 649- 72 & 10.472 & 0.633  & 1.006 &2000 Sep 30 & KP 2.1m & M6.5\tablenotemark{a} & 10.5$\pm$0.3 & 10.1$\pm$1.4  & 1     \\
0334106$-$213034 &                     & 11.957 & 0.706  & 1.121 &2002 Jan 27 & CT 1.5m & M6                    & 10.1$\pm$0.4 & 23.3$\pm$4.0  &       \\
0354200$-$143738 &                     & 11.358 & 0.602  & 1.033 &2002 Jan 28 & CT 1.5m & M6.5                  & 10.5$\pm$0.3 & 15.2$\pm$2.1  &       \\
0410480$-$125142 & LP 714- 37          & 11.060 & 0.591  & 1.045 &2000 Sep 30 & KP 2.1m & M6\tablenotemark{b}   & 10.1$\pm$0.4 & 15.4$\pm$2.7  & 1,2,3 \\
1124532$+$132253 &                     & 11.042 & 0.643  & 1.014 &2002 Jan 24 & KP 4m   & M5.5                  & \phn9.6$\pm$0.2 & 19.8$\pm$1.9  &       \\
1424187$-$351432 &                     & 11.809 & 0.625  & 1.019 &2002 Jan 27 & CT 1.5m & M6.5                  & 10.5$\pm$0.3 & 18.6$\pm$2.6  &       \\
1431304$+$171758 &                     & 12.217 & 0.649  & 1.054 &2002 Jan 24 & KP 4m   & M6.5                  & 10.5$\pm$0.3 & 22.5$\pm$3.2  &       \\
\enddata

\tablenotetext{a}{ST change from M5.5 in Paper III}
\tablenotetext{b}{ST change from M5.5 in Paper III}
\tablenotetext{c}{ST change from M5.5 in Paper III}

\tablerefs{(1) Paper III; (2) \citet{wilson}; (3) \citet{NN5}}

\end{deluxetable}

\pagestyle{empty}
\begin{deluxetable}{llllllllllc}
\tablewidth{0pt} \tabletypesize{\scriptsize} \rotate
\tablecaption{Early-type M dwarfs Discovered Outside 20
pc\label{tab:earlyfar}} \tablehead{ \colhead{2MASS} &
\colhead{Other} & \colhead{$J$} & \colhead{$J-H$} &
\colhead{$J-K_S$} & \colhead{Obs.} & \colhead{Telescope} &
\colhead{Spectral}& \colhead{$M_J$}& \colhead{$d$}& \colhead{Other}\\
\colhead{Designation} & \colhead{Names} & \colhead{} & \colhead{}
& \colhead{} & \colhead{Date} & \colhead{} & \colhead{Type}&
\colhead{}& \colhead{(pc)}& \colhead{Ref.} } \startdata
0014083$+$493433 &                               & 13.276 & 0.621  & 1.022 &2001 Jul 15 & KP 4m   & M6                    & 10.1$\pm$0.4 & 42.8$\pm$7.4  &     \\
0017533$+$145324 &                               & 13.204 & 0.690  & 1.056 &2001 Jul 16 & KP 4m   & M6.5                  & 10.5$\pm$0.3 & 35.5$\pm$5.0  & 1   \\
0030198$+$314125 &                               & 13.383 & 0.630  & 1.006 &2001 Jul 21 & KP 4m   & M4                    &  8.9$\pm$0.2 & 77.4$\pm$6.2  &     \\
0104081$+$362301 &                               & 13.064 & 0.637  & 1.015 &2001 Jul 23 & KP 4m   & M5                    &  9.4$\pm$9.3 & 54.8$\pm$7.1  &     \\
0219331$+$141632 &                               & 12.725 & 0.660  & 1.014 &2001 Jul 23 & KP 4m   & M4                    &  7.6$\pm$0.3 &  104$\pm$13   &     \\
0234295$+$361311 &                               & 12.459 & 0.637  & 1.002 &2000 Sep 30 & KP 2.1m & M6\tablenotemark{a}   & 10.1$\pm$0.4 & 29.4$\pm$5.1  & 2   \\
0250023$-$080841 & LP 651- 17/LHS 1450           & 11.878 & 0.652  & 1.028 &2000 Sep 30 & KP 2.1m & M5.5                  &  9.7$\pm$0.2 & 26.9$\pm$2.7  & 2   \\
0301032$+$441656 &                               & 12.059 & 0.681  & 1.061 &2000 Oct 02 & KP 2.1m & M6                    & 10.1$\pm$0.4 & 24.5$\pm$4.2  &     \\
\enddata

\tablenotetext{a}{ST change from M5 in Paper III}
\tablenotetext{b}{ST change from M5.5 in Paper III}
\tablenotetext{c}{ST change from M5.5 in Paper III}
\tablerefs{(1) \citet{wilson}; (2) Paper III; (3) \citet{NN4}; (4) \citet{NN5}}
\end{deluxetable}

\pagestyle{empty}
\begin{deluxetable}{llllllr}
\tablewidth{0pt}
\tabletypesize{\small}
\rotate
\tablecaption{Young Objects\label{tab:young}}
\tablehead{\colhead{2MASSI} & \colhead{$J$} & \colhead{$J-H$} &
\colhead{$J-K_S$} & \colhead{Obs.} & \colhead{Telescope} & \colhead{Comments} \\
\colhead{Designation} & \colhead{} & \colhead{} &
\colhead{} & \colhead{date} & \colhead{} & \colhead{}}
\startdata
0253597$+$320637 & 13.623 & 0.693 & 1.079 & 2002 Jan 25 & KP 4m   & \\
0435145$-$141446 & 11.868 & 1.236 & 1.927 & 2002 Jan 24 & CT 4m   &  \\
0608528$-$275358 & 13.600 & 0.702 & 1.210 & 2002 Jan 25 & CT 4m   & enhanced VO absorption \\
0619526$-$290359 & 15.101 & 0.882 & 1.649 & 2002 Jan 24 & CT 4m   & \\
2234416$+$404138 & 12.569 & 0.735 & 1.138 & 2000 Oct 02 & KP 2.1m & strong H$\alpha$ emission \\
\enddata
\end{deluxetable}

\pagestyle{empty}
\begin{deluxetable}{lllllllr}
\tablewidth{0pt} \tabletypesize{\small} \rotate
\tablecaption{Spectroscopically Confirmed
Giants\label{tab:giants}} \tablehead{\colhead{2MASSI} &
\colhead{$J$} & \colhead{$J-H$} &
\colhead{$J-K_S$} & \colhead{Obs.} & \colhead{Telescope} & \colhead{Spectral} & \colhead{Comments} \\
\colhead{Designation} & \colhead{} & \colhead{} &
\colhead{} & \colhead{date} & \colhead{} & \colhead{Type} & \colhead{}}
\startdata
0012342$-$225516 &    10.393  & 0.779 & 1.125 & 2002 Jan 31 & CT 1.5m & M1 III     & \\
0028097$+$121719 &    10.177  & 0.883 & 1.392 & 2000 Oct 02 & KP 2.1m & M5 III     & \\
0036441$-$060858 &\phn9.998   & 0.785 & 1.139 & 2002 Jan 28 & CT 1.5m & M1 III     & \\
0229220$-$110612 &    13.213  & 0.687 & 1.325 & 2002 Jan 31 & CT 1.5m & M0 III     & \\
0252543$-$114220 &    12.505  & 0.813 & 1.171 & 2002 Jan 30 & CT 1.5m & M1 III     & \\
0424503$-$740443 &    12.337  & 0.760 & 1.162 & 2002 Jan 26 & CT 4m   & M8 III     & \\
0436168$-$641131 &    12.098  & 0.863 & 1.275 & 2002 Jan 26 & CT 4m   & M5 III     & \\
0534324$-$121631 &    10.161  & 0.721 & 1.078 & 2002 Jan 27 & CT 1.5m & M0 III     & \\
\enddata
\end{deluxetable}

\pagestyle{empty}
\begin{deluxetable}{lllllllr}
\tabletypesize{\small} \tablewidth{0pt}
\tablecaption{Spectroscopically Confirmed Carbon
Stars\label{tab:carbon}} \tablehead{ \colhead{2MASSI} &
\colhead{$J$} & \colhead{$J-H$} & \colhead{$J-K_S$} &
\colhead{Obs.} & \colhead{Telescope} & \colhead{Visual
pm\tablenotemark{a}} &
\colhead{Comments}\\
\colhead{Designation} & \colhead{} & \colhead{} & \colhead{} & \colhead{
Date} & \colhead{} & \colhead{(epochs)} &
\colhead{}}
\startdata
0236246$-$204103 & 13.673   & 0.768 & 1.224 & 2002 Jan 26 & CT 4m   & yes (12/79--11/98) & dC, LP 830-18\tablenotemark{b} \\
0402332$-$681623 & 13.490   & 1.420 & 2.372 & 2002 Jan 31 & CT 1.5m & no (12/92--01/00)  &                                \\
0426387$+$142516 & 13.227   & 1.211 & 2.014 & 2000 Oct 02 & KP 2.1m & no (12/55--11/98)  &                                \\
0431342$-$723208 & 13.185   & 1.234 & 1.958 & 2002 Jan 31 & CT 1.5m & no (01/93--12/98)  &                                \\
0459495$-$750918 & 12.638   & 1.327 & 2.325 & 2002 Jan 26 & CT 4m   & no (01/75--12/98)  &                                \\
0614383$-$135137 &\phn9.829 & 0.789 & 1.127 & 2002 Jan 27 & CT 1.5m & no (11/83--01/99)  &                                \\
0635542$-$645226 & 12.826   & 0.945 & 1.395 & 2002 Jan 23 & CT 4m   & no (11/82--02/00)  &                                \\
0654386$-$713059 & 13.060   & 1.244 & 2.083 & 2002 Jan 30 & CT 1.5m & no (02/91--12/98)  &                                \\
\enddata

\tablenotetext{a}{Possible visual proper motion determined by PJL from
examining DSS, XDSS and 2MASS images.  Epochs listed are either DSS--2MASS or XDSS $R$--2MASS.}

\tablenotetext{b}{A discussion of these new carbon dwarfs will appear in P.J. Lowrance et
al., in preparation.}
\end{deluxetable}

\pagestyle{headings}
\begin{deluxetable}{llrc}
\tablewidth{0pt}
\tablecaption{$M_J$/Spectral Type Calibration Data\label{tab:mjcal}} \tablehead{
\colhead{Name} & \colhead{Spectral} & \colhead {$M_J$} &
\colhead {Ref.} \\
\colhead{} & \colhead{Type\tablenotemark{a}} & \colhead {} &
\colhead {}
}
\startdata
Gl 551                    &   5.5  &   9.68$\pm$0.01     &   1   \\
GJ 1286                   &   5.5  &   9.90$\pm$0.05     &   1   \\
GJ 1002                   &   5.5  &   9.98$\pm$0.03     &   1   \\
LHS 3339                  &   5.5  &   10.05$\pm$0.06    &   3   \\
Gl 412B                   &   6    &   10.04$\pm$0.01    &   1   \\
LHS 1443                  &   6    &   10.09$\pm$0.20    &   2   \\
LHS 1516                  &   6    &   10.11$\pm$0.26    &   2   \\
Gl 406                    &   6    &   10.15$\pm$0.01    &   1   \\
\enddata
\tablenotetext{a}{M5 = 5, L0 = 10, L8 = 18}
\tablerefs{(1) 8 pc Sample; (2) PMSU; (3) \citet{Dahn}}
\end{deluxetable} 
\begin{deluxetable}{lrllllr}
\tabletypesize{\scriptsize}
\tablewidth{0pt} \tablecaption{$J<9$ Sources with IRAS Catalog
Counterparts \label{tab:iras}} \tablehead{\colhead{$\alpha$
(2000)} & \colhead {$\delta$} & \colhead{$J-H$} & \colhead
{$H-K_S$} & \colhead {$K_S$ } & \colhead {$J-K_S$ } &  \colhead{ID
}} \startdata
00 01 21.4 &    51 12 13 & 0.920   &   0.700 &   6.910 & 1.620   & IRAS 23587+5055        \\
00 20 53.4 &    50 09 07 & 0.930   &   0.590 &   5.480 & 1.520   & IRAS 00182+4952        \\
00 24 18.7 &    34 23 04 & \nodata & \nodata &   4.830 & 1.060   & IRAS 00216+3406, M0    \\
01 49 43.1 &    38 40 53 & \nodata & \nodata &   4.960 & 1.100   & IRAS 01467+3826        \\
02 39 05.1 &    31 00 26 & 0.850   &   0.450 &   6.070 & 1.300   & IRAS 02361+3047        \\
03 03 52.0 &    45 11 42 & 0.950   &   0.530 &   5.360 & 1.480   & IRAS 03005+4459        \\
03 06 47.1 &    28 20 32 & 6.300   & \nodata &   4.890 & 1.410   & IRAS 03037+2808        \\
03 15 24.2 &    40 44 54 & 0.800   &   0.320 &   5.350 & 1.120   & IRAS 03121+4033        \\
\enddata
\end{deluxetable}

\begin{deluxetable}{lrlllllr}
\tabletypesize{\scriptsize}
\tablewidth{0pt} \tablecaption{$J<9$ Sources with Stellar
Counterparts\label{tab:stellar}} \tablehead{ \colhead{$\alpha$
(2000)} & \colhead {$\delta$} & \colhead {$V$} & \colhead{$J-H$} &
\colhead {$H-K_S$ } & \colhead {$K_S$ } & \colhead {$J-K_S$ } &
\colhead{ID }} \startdata
00 20 01.3 & 51 53 26 &    \phn8.72&   5.630 &  \nodata&  \nodata& \nodata& HD 232159, K5     \\
00 38 06.2 & 37 35 21 &   10.4 &   0.720 &   0.340 &   5.900 & 1.060 & GSC 02784-01363, M     \\
00 42 32.4 & 33 11 31 &    \phn9.57&   0.740 &   0.300 &   5.480 & 1.040 & BD+32 116, K5     \\
00 44 41.2 & 47 55 47 &    \phn9.61&   0.670 &   0.490 &   5.920 & 1.160 & BD+47 182, K5     \\
01 15 51.7 &$-$20 34 59&   10.1 &   0.689&   0.325&   6.185& 1.014& BD-21 199                        \\
01 28 55.4 & 34 40 19 &   10.4 &   0.690 &   0.310 &   6.220 & 1.000 &  GSC 02300-01336    \\
01 48 29.8 & 40 16 31 &   10.02&   0.690 &   0.350 &   5.650 & 1.040 &  GSC 02819-01779     \\
02 04 31.5 & 15 37 48 &    \phn9.65&   5.470 & \nodata & \nodata & \nodata &  BD+14 334, M2    \\
\enddata
\end{deluxetable}

\begin{deluxetable}{lrlllllr}
\tabletypesize{\scriptsize}
\tablewidth{0pt} \tablecaption{Carbon
stars\label{tab:brightcarbon}} \tablehead{\colhead{$\alpha$
(2000)} & \colhead {$\delta$} & \colhead {$V$} & \colhead{$J-H$} &
\colhead {$H-K_S$ } & \colhead {$K_S$ } & \colhead {$J-K_S$ } &
\colhead{ID }} \startdata
01 58 51.1 &    40 37 20 &   12.1 &   1.180 &   0.740 &   6.630 & 1.920 & CGCS 298   \\
02 13 29.9 &    46 20 20 &   13.5 &   0.860 &   0.480 &   6.570 & 1.340 & CGCS 322    \\
03 08 48.7 &    27 54 59 & \nodata&   1.030 &   0.530 &   6.190 & 1.560 & CGCS 444    \\
03 12 46.9 &    38 47 50 & \nodata&   0.970 &   0.580 &   6.070 & 1.550 & V458 Per    \\
03 27 31.3 &    39 04 50 & \nodata&   1.110 &   0.530 &   7.020 & 1.640 & FBS 0324+389 \\
06 14 35.9 & $-$15 01 09 &   13.0 &   1.034 &   0.565 &   6.382 & 1.599 & CGCS 1210      \\
06 15 27.2 &    47 54 25 & \nodata&   1.232 &   0.838 &   5.883 & 2.070 & CGCS 1194       \\
07 00 01.7 &    35 05 56 & \nodata&   1.126 &   0.723 &   6.628 & 1.849 & FBS 0656+351    \\
07 06 07.7 &    40 12 01 & \nodata&   1.076 &   0.523 &   7.088 & 1.599 & FBS 0702+402    \\
07 54 13.2 &    72 13 39 &   13.8 &   1.050 &   0.520 &   5.985 & 1.570 & CGCS 1876       \\
11 45 01.4 &    64 47 37 & \nodata&   0.777 &   0.489 &   6.034 & 1.266 & GSC 04156-00392 \\
12 24 32.7 & $-$26 14 08 &   13.6 &   0.815 &   0.369 &   7.219 & 1.184 & KV Hya          \\
20 03 34.2 & $-$25 27 15 &   12.0 &   0.783 &   0.319 &   7.261 & 1.102 & CD-25 14520     \\
\enddata
\end{deluxetable}

\begin{deluxetable}{lrrllllr}
\tabletypesize{\scriptsize} \tablewidth{0pt} \tablecaption{Miras
and Long-period Variables\label{tab:miras}} \tablehead{
\colhead{$\alpha$ (2000)} & \colhead {$\delta$} & \colhead {$V$} &
\colhead{$J-H$} & \colhead {$H-K_S$ } & \colhead {$K_S$ } &
\colhead {$J-K_S$ } &  \colhead{ID }} \startdata
00 09 36.8 &   37 47 31 &   12.4 &   1.010 &  0.530 &   5.430 & 1.540  & V414 And     \\
00 38 52.9 &   45 33 49 & \nodata&   0.780 &  0.490 &   7.710 & 1.270  & V403 And     \\
00 59 46.4 &   27 56 44 &   11.4 &   0.770 &  0.430 &   5.980 & 1.200  & W Psc, M2e     \\
01 22 52.9 &   25 23 06 & \nodata& \nodata & \nodata&   4.580 & 1.400  & TZ Psc      \\
01 58 06.1 &   38 39 18 & \nodata&   0.800 &  0.360 &   6.830 & 1.160  &  OY And    \\
02 59 56.2 &   42 37 15 & \nodata&   0.850 &  0.630 &   5.720 & 1.480  & IV Per     \\
03 11 40.7 &   41 38 46 & \nodata&   0.840 &  0.450 &   5.210 & 1.290  &  GG Per    \\
03 18 27.8 &   42 49 18 & \nodata&   0.820 &  0.360 &   6.230 & 1.180  &  WW Per    \\
\enddata
\end{deluxetable}

\begin{deluxetable}{lrrllllr}
\tabletypesize{\scriptsize} \tablewidth{0pt}
\tablecaption{Semi-regular Variables\label{tab:var}} \tablehead{
\colhead{$\alpha$ (2000)} & \colhead
{$\delta$} & \colhead {$V$} & \colhead{$J-H$} & \colhead {$H-K_S$
} & \colhead {$K_S$ } & \colhead {$J-K_S$ } &  \colhead{ID }}
\startdata
00 45 01.1 &   48 41 02 &   10.7 &   1.250 &   0.720 &   5.300 & 1.970 & V864 Cas    \\
00 46 24.8 &   47 41 33 & \nodata&   1.280 &   0.970 &   4.990 & 2.250 &  GSC 03266-01510  \\
02 34 31.4 &   40 04 06 & \nodata&   0.830 &   0.450 &   7.330 & 1.280 & GU And     \\
03 21 10.4 &$-$17 13 56 &   10.0 &   0.693 &   0.371 &   6.960 & 1.064 & CO Eri      \\
08 22 01.4 &   13 37 05 &   10.8 &   0.799 &   0.362 &   5.833 & 1.161 & GT Cnc       \\
08 27 40.4 &   19 15 43 &   11.4 &   0.730 &   0.369 &   6.718 & 1.099 & GV Cnc      \\
08 28 08.0 &   38 20 22 &   11.1 &   0.780 &   0.353 &   6.962 & 1.133 & RX Lyn      \\
09 06 39.0 &$-$19 18 44 &   12.7 &   1.238 &   0.843 &   6.009 & 2.081 & V379 Hya    \\
15 06 42.1 &$-$23 25 38 &   12.0 &   0.761 &   0.520 &   6.082 & 1.281 & WZ Lib,     \\
16 23 13.1 &   44 08 28 &   10.5 &   0.786 &   0.318 &   7.623 & 1.104 & AY Her      \\
19 22 42.9 &$-$32 08 03 &    9.6 &   0.733 &   0.322 &   5.987 & 1.055 & AM Sgr      \\
19 32 44.7 &$-$32 14 14 &   12.5 &   0.839 &   0.448 &   5.047 & 1.287 & V347 Sgr    \\
20 51 00.6 &   14 31 15 &   12.0 &   0.800 &   0.335 &   7.616 & 1.135 & DZ Del    \\
21 15 02.8 &$-$09 46 37 & \nodata&   0.876 &   0.491 &   5.907 & 1.367 & BY Aqr      \\
23 21 05.5 &   45 24 30 & \nodata&   0.750 &   0.320 &   7.520 & 1.070 & V338 And     \\
23 25 41.7 &   45 42 04 & \nodata&   1.480 &   0.960 &   6.280 & 2.440 & V339 And     \\
\enddata
\end{deluxetable}

\begin{deluxetable}{lrrllllr}
\tabletypesize{\scriptsize} \tablewidth{0pt} \tablecaption{Other
Late-type Stars\label{tab:other}}
\tablehead{\colhead{$\alpha$ (2000)} & \colhead {$\delta$} & \colhead {$V$}
& \colhead{$J-H$} & \colhead {$H-K_S$ } & \colhead {$K_S$ } &
\colhead {$J-K_S$ } &  \colhead{ID}} \startdata
04 27 42.7 &$-$73 11 08 & \nodata&   0.795&   0.316&   7.742& 1.111& WOH S 11, M                       \\
08 53 26.7 &$-$13 31 16 & \nodata&   0.802&   0.374&   6.566& 1.176& V357 Hya, M-type variable        \\
09 31 01.7 &$-$23 09 00 &   11.0 &   0.686&   0.319&   7.526& 1.005& AT Hya Irr                     \\
09 56 32.9 &$-$10 01 17 &   15.0 &   0.846&   0.660&   5.755& 1.506& BR B0954-0947, M8 III             \\
12 44 42.0 &   36 45 50 &    9.8 &   0.797&   0.420&   6.198& 1.217& TX CVn symbiotic                 \\
15 01 12.8 &$-$25 39 24 &   11.6 &   0.917&   0.493&   7.126& 1.410& BV 1671, variable                     \\
15 12 15.2 &$-$22 59 07 &   12.0 &   0.856&   0.438&   5.690& 1.294& StM 218, M6                       \\
17 25 13.0 &   64 26 12 & \nodata&   0.867&   0.471&   6.603& 1.338& FBS 1724+644, M                   \\
\enddata
\end{deluxetable}

\begin{deluxetable}{lrllllr}
\tabletypesize{\scriptsize} \tablewidth{0pt} \tablecaption{$J<9$
Sources Without a Cataloged Counterpart\label{tab:unmatched}}
\tablehead{\colhead{$\alpha$ (2000)} & \colhead {$\delta$}
 & \colhead{$J-H$ } & \colhead {$H-K_S$ } & \colhead {$K_S$ } &
\colhead {$J-K_S$ } &  \colhead{$R_{USNO}$ }}
\startdata
00 11 11.0 &   48 58 43 & 0.800 &  0.320 &   5.750 & 1.120 & 10.9     \\
00 12 10.5 &   46 53 30 & 0.800 &  0.310 &   6.700 & 1.110 & 10.8     \\
00 14 40.9 &   40 34 52 & 0.720 &  0.460 &   6.350 & 1.180 & 10.2     \\
00 19 29.8 &   48 38 27 & 0.790 &  0.310 &   7.800 & 1.100 & 11.8     \\
00 21 48.9 &   44 52 18 & 0.800 &  0.420 &   7.490 & 1.220 &  9.1     \\
00 28 26.9 &   31 24 11 & 0.800 &  0.310 &   6.630 & 1.110 &  9.9     \\
00 38 59.3 &$-$19 27 42 & 0.788 &  0.355 &   6.733 & 1.143 & 10.9     \\
00 57 50.8 &   50 50 07 & 0.900 &  0.450 &   7.370 & 1.350 & 13.9     \\
\enddata
\end{deluxetable}


\end{document}